\documentclass[english,superscriptaddress]{revtex4-1}
\usepackage[T1]{fontenc}
\usepackage[latin9]{inputenc}
\usepackage{geometry}
\usepackage{amsmath}
\usepackage{amssymb}
\usepackage{graphicx}

\makeatletter

\providecommand{\tabularnewline}{\\}

\makeatother

\usepackage{babel}
\begin{document}
\title{Relativistic spin dynamics for vector mesons}
\author{Xin-Li Sheng}
\affiliation{Key Laboratory of Quark and Lepton Physics (MOE) and Institute of
Particle Physics, Central China Normal University, Wuhan, 430079,China}
\author{Lucia Oliva}
\affiliation{Department of Physics and Astronomy \textquotedbl Ettore Majorana\textquotedbl ,
University of Catania, Via S. Sofia 64, I-95123 Catania, Italy}
\affiliation{INFN Sezione di Catania, Via S. Sofia 64, I-95123 Catania, Italy}
\author{Zuo-Tang Liang}
\affiliation{Key Laboratory of Particle Physics and Particle Irradiation (MOE),
Institute of Frontier and Interdisciplinary Science, Shandong University,
Qingdao, Shandong 266237, China}
\author{Qun Wang}
\affiliation{Peng Huanwu Center for Fundamental Theory and Department of Modern
Physics, University of Science and Technology of China, Hefei, Anhui
230026, China}
\author{Xin-Nian Wang}
\affiliation{Nuclear Science Division, MS 70R0319, Lawrence Berkeley National Laboratory,
Berkeley, California 94720, USA}
\begin{abstract}
We propose a relativistic theory for spin density matrices of vector
mesons based on Kadanoff-Baym equations in the closed-time-path formalism.
The theory puts the calculation of spin observables such as the spin
density matrix element $\rho_{00}$ for vector mesons on a solid ground.
Within the theory we formulate $\rho_{00}$ for $\phi$ mesons into
a factorization form in separation of momentum and space-time variables.
We argue that the main contribution to $\rho_{00}$ at lower energies
should be from the $\phi$ fields that can polarize the strange quark
and antiquark in the same way as electromagnetic fields. The key observation
is that there is correlation inside the $\phi$ meson wave function
between the $\phi$ field that polarizes the strange quark and that
polarizes the strange antiquark. This is reflected by the fact that
the contributions to $\rho_{00}$ are all in squares of fields which
are nonvanishing even if the fields may strongly fluctuate in space-time.
The fluctuation of strong force fields can be extracted from $\rho_{00}$
of quarkonium vector mesons as links to fundamental properties of
quantum chromodynamics.
\end{abstract}
\maketitle

\section{Introduction}

There is an intrinsic connection between rotation and spin polarization
since they are related to the conservation of total angular momentum
and can be converted from one to another, as demonstrated in the Barnett
effect \citep{Barnett:1935} and the Einstein-de Haas effect \citep{dehaas:1915}
in materials. A recent example is the observation of a spin-current
from the vortical motion in a liquid metal \citep{Takahashi:2016}.
The same effects also exist in high energy heavy-ion collisions (HIC)
in which the huge orbital angular momentum (OAM) along the direction
normal to the reaction plane can be partially converted to the global
spin polarization of hadrons \citep{Liang:2004ph,Liang:2004xn,Voloshin:2004ha,Betz:2007kg,Becattini:2007sr,Gao:2007bc}
(see, e.g. \citep{Wang:2017jpl,Florkowski:2018fap,Becattini:2020ngo,Gao:2020lxh,Huang:2020dtn},
for recent reviews). The global spin polarization of $\Lambda$ (including
$\overline{\Lambda}$ hereafter) has been measured through their weak
decays in Au+Au collisions at $\sqrt{s_{NN}}=7.7-200$ GeV \citep{STAR:2017ckg,Adam:2018ivw}.

As spin-one particles, vector mesons can also be polarized in heavy
ion collisions in the same way as hyperons. Normally the spin states
of vector mesons are described by the spin density matrix element
$\rho_{\lambda_{1}\lambda_{2}}$ with $\lambda_{1},\lambda_{2}=0,\pm1$
labeling spin states along the spin quantization direction. The vector
mesons mainly decay through strong interaction that respects parity
symmetry. So their spin polarization proportional to $\rho_{11}-\rho_{-1,-1}$
cannot be measured through their decays. Instead, $\rho_{00}$ can
be measured through the angular distribution of its decay daughters
\citep{Liang:2004xn,Yang:2017sdk,Tang:2018qtu,Goncalves:2021ziy}.
If $\rho_{00}=1/3$, the spin states are equally populated in the
three spin states implying that the vector meson is not polarized.
If $\rho_{00}\neq1/3$, the three spin states are not equally populated,
so the spin of the vector meson is aligned either in the direction
of the spin quantization or of the transverse direction perpendicular
to it. In 2008, the STAR collaboration measured $\rho_{00}$ for the
vector meson $\phi(1020)$ in Au+Au collisions at 200 GeV, but the
result is consistent with $1/3$ within errors due to statistics \citep{Abelev:2008ag}.
Recently STAR has measured the $\phi$ meson's $\rho_{00}$ at lower
energies which shows a significant deviation from $1/3$ \citep{STAR:2022fan}.
It can hardly be explained by conventional mechanism \citep{Yang:2017sdk,Xia:2020tyd,Gao:2021rom,Muller:2021hpe}.
In Ref. \citep{Sheng:2019kmk}, some of us proposed that a large deviation
of $\rho_{00}$ from 1/3 for the $\phi$ meson may possibly arise
from the $\phi$ field, a strong force field with vacuum quantum number
in connection with pseudo-Goldstone bosons and vacuum properties of
quantum chromodynamics. Such a proposal is based on a nonrelativistic
quark coalescence model for the spin density matrix of vector mesons
\citep{Yang:2017sdk,Sheng:2020ghv}.

In this paper we will present a relativistic theory for the spin density
matrix of vector mesons from the Kadanoff-Baym (KB) equation \citep{Kadanoff:1962}
in the closed-time-path (CTP) formalism \citep{Martin:1959jp,Keldysh:1964ud}
(for reviews of the KB equation and the CTP formalism, we refer the
readers to Refs. \citep{Chou:1984es,Blaizot:2001nr,Berges:2004yj,Cassing:2008nn}).
Then we can derive the spin Boltzmann equation for vector mesons with
their spin density matrices being expressed in terms of the matrix
valued spin dependent distributions (MVSD) of the quarks and antiquarks
\citep{Sheng:2021kfc}. This puts the calculation of spin observables
such as $\rho_{00}$ for vector mesons on a solid ground.

The paper is organized as follows. In Sec. \ref{sec:green-function}
we will give an introduction to Green functions on the CTP for vector
mesons which can be expressed in MVSD. In Sec. \ref{sec:kb-vector}
the KB equations for vector mesons are derived. In Sec. \ref{sec:spin-density-matrix}
the spin density matrices for vector mesons will be formulated from
the spin Boltzmann equations. In Sec. \ref{sec:spin-density-phi}
the spin density matrices for $\phi$ mesons will be evaluated. Discussions
on the main results and conclusions are given in the final section,
Sec. \ref{sec:discussions}.

We adopt the sign convention for the metric tensor $g^{\mu\nu}=g_{\mu\nu}=\mathrm{diag}(1,-1,-1,-1)$
where $\mu,\nu=0,1,2,3$. The sign convention for the Levi-Civita
symbol is $\epsilon^{0123}=-\epsilon_{0123}=1$. We can write the
space-time coordinate as $x=x^{\mu}=(x^{0},\mathbf{x})=(t,\mathbf{x})$
and $x_{\mu}=(x_{0},-\mathbf{x})$ with $x_{0}=x^{0}=t$. The four-momentum
for a particle is denoted as $p=p^{\mu}=(p^{0},\mathbf{p})$ or $p_{\mu}=(p_{0},-\mathbf{p})$,
if it is on-shell we have $p_{0}=p^{0}=\sqrt{\mathbf{p}^{2}+m^{2}}\equiv E_{p}=E_{\mathbf{p}}$.
Normally we use Greek letters to denote four-dimensional indices of
four-vectors and four-tensors and Latin letters to denote their spatial
components.

\section{Green functions on CTP for vector mesons}

\label{sec:green-function}The massive spin-1 particle, such as the
vector meson with the mass $m_{V}$, can be described by the vector
field $A_{V}^{\mu}(x)$ with the classical Lagrangian density 
\begin{equation}
\mathcal{L}=-\frac{1}{4}F_{\mu\nu}^{V}F_{V}^{\mu\nu}+\frac{m_{V}^{2}}{2}A_{\mu}^{V}A_{V}^{\mu}-A_{\mu}^{V}j^{\mu}.\label{eq:lag-vector}
\end{equation}
where $j^{\mu}$ is the source current, $F_{V}^{\mu\nu}=\partial^{\mu}A_{V}^{\nu}-\partial^{\nu}A_{V}^{\mu}$
is the field strength tensor, and $A_{V}^{\mu}$ is assumed to be
the real classical field for the charge (including flavor) neutral
particle. From $\mathcal{L}$ one can obtain the Proca equation \citep{Proca:1936,Itzykson:1980rh}
\begin{equation}
L^{\mu\nu}(x)A_{\nu}^{V}(x)=j^{\mu}(x),
\end{equation}
where the differential operator is defined as 
\begin{equation}
L^{\mu\nu}(x)=\left(\partial_{x}^{2}+m_{V}^{2}\right)g^{\mu\nu}-\partial_{x}^{\mu}\partial_{x}^{\nu}.
\end{equation}
A constraint equation can be derived by contracting the above equation
with $\partial_{\mu}$ as 
\begin{equation}
\partial_{\mu}A_{V}^{\mu}(x)=\frac{1}{m_{V}^{2}}\partial_{\mu}j^{\mu}(x)=0,\label{eq:constraint}
\end{equation}
if the source current is conserved $\partial_{\mu}j^{\mu}=0$. The
above equation means that the longitudinal component of $A_{V}^{\mu}(x)$
is vanishing for the conserved current.

The free vector field can be quantized as 
\begin{eqnarray}
A_{V}^{\mu}(x) & = & \sum_{\lambda=0,\pm1}\int\frac{d^{3}k}{(2\pi\hbar)^{3}}\frac{1}{2E_{k}^{V}}\nonumber \\
 &  & \times\left[\epsilon^{\mu}(\lambda,{\bf k})a_{V}(\lambda,{\bf k})e^{-ik\cdot x/\hbar}+\epsilon^{\mu\ast}(\lambda,{\bf k})a_{V}^{\dagger}(\lambda,{\bf k})e^{ik\cdot x/\hbar}\right],\label{eq:a-quantization}
\end{eqnarray}
where $E_{k}^{V}=\sqrt{\mathbf{k}^{2}+m_{V}^{2}}$ and $\lambda$
denote the energy and the spin state in the spin quantization direction
respectively, the creation and annihilation operators $a_{V}(\lambda,{\bf k})$
and $a_{V}^{\dagger}(\lambda,{\bf k})$ satisfy the commutator
\begin{equation}
\left[a_{V}(\lambda,{\bf k}),a_{V}^{\dagger}(\lambda^{\prime},{\bf k}^{\prime})\right]=\delta_{\lambda\lambda^{\prime}}2E_{{\bf k}}(2\pi\hbar)^{3}\delta^{(3)}({\bf k}-{\bf k}^{\prime}),
\end{equation}
and the polarization vector $\epsilon^{\mu}(\lambda,{\bf k})$ obeys
\begin{align}
 & k_{\mu}\epsilon^{\mu}(\lambda,{\bf k})=0,\nonumber \\
 & \epsilon(\lambda,{\bf k})\cdot\epsilon^{*}(\lambda^{\prime},{\bf k})=-\delta_{\lambda\lambda^{\prime}},\nonumber \\
 & \sum_{\lambda}\epsilon^{\mu}(\lambda,{\bf k})\epsilon^{\nu*}(\lambda,{\bf k})=-\left(g^{\mu\nu}-\frac{k^{\mu}k^{\nu}}{m_{V}^{2}}\right).\label{eq:polar-vector-rel}
\end{align}
In above relations, the first one follows the constraint (\ref{eq:constraint})
and $k^{\mu}=(E_{k}^{V},\mathbf{k})$ denotes the on-shell four-momentum
for the vector meson. By the field quantization in (\ref{eq:a-quantization}),
one can check that $A_{V}^{\mu}(x)$ is Hermitian, i.e. $A_{V}^{\mu\dagger}(x)=A_{V}^{\mu}(x)$.

\begin{figure}[h]
\includegraphics[scale=0.6]{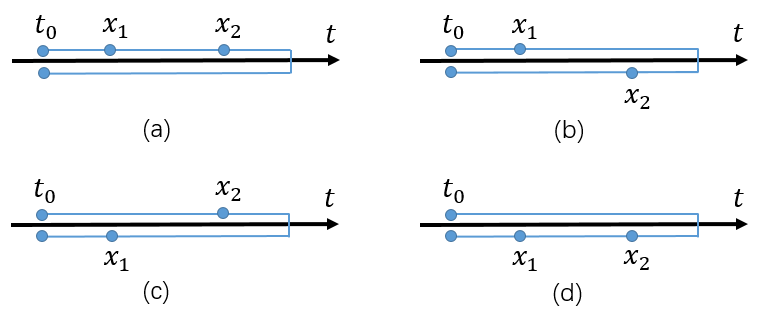}

\caption{\label{fig:ctp}The closed-time path and four components of the two-point
Green function on CTP. The positive and negative time-branches are
denoted as $C_{+}$ and $C_{-}$ respectively. (a) $x_{1}^{0}=t_{1}\in C_{+}$,
$x_{2}^{0}=t_{2}\in C_{+}$; (b) $x_{1}^{0}=t_{1}\in C_{+}$, $x_{2}^{0}=t_{2}\in C_{-}$;
(c) $x_{1}^{0}=t_{1}\in C_{-}$, $x_{2}^{0}=t_{2}\in C_{+}$; (d)
$x_{1}^{0}=t_{1}\in C_{-}$, $x_{2}^{0}=t_{2}\in C_{-}$.}
\end{figure}

One can define the two-point Green function for the vector meson on
the CTP 
\begin{equation}
G_{\mathrm{CTP}}^{\mu\nu}(x_{1},x_{2})\equiv\left\langle T_{C}A_{V}^{\mu}(x_{1})A_{V}^{\nu\dagger}(x_{2})\right\rangle ,
\end{equation}
where $x_{1}$ and $x_{2}$ are two space-time points whose time components
are defined on the CTP contour and $T_{C}$ represents the time-ordering
on the CTP contour. We can write $G_{\mu\nu}^{\mathrm{CTP}}(x_{1},x_{2})$
in a matrix form 
\begin{equation}
G_{\mu\nu}^{\mathrm{CTP}}(x_{1},x_{2})=\left(\begin{array}{cc}
G_{\mu\nu}^{++}(x_{1},x_{2}) & G_{\mu\nu}^{+-}(x_{1},x_{2})\\
G_{\mu\nu}^{-+}(x_{1},x_{2}) & G_{\mu\nu}^{--}(x_{1},x_{2})
\end{array}\right)=\left(\begin{array}{cc}
G_{\mu\nu}^{F}(x_{1},x_{2}) & G_{\mu\nu}^{<}(x_{1},x_{2})\\
G_{\mu\nu}^{>}(x_{1},x_{2}) & G_{\mu\nu}^{\overline{F}}(x_{1},x_{2})
\end{array}\right).\label{eq:green-func}
\end{equation}
The '$++$' component of $G_{\mu\nu}^{\mathrm{CTP}}$ with both $t_{1}$
and $t_{2}$ (time components of $x_{1}$ and $x_{2}$) on the positive
time-branch is just the Feynman propagator $G_{\mu\nu}^{F}(x_{1},x_{2})$
as shown in Fig. \ref{fig:ctp}(a). The '$+-$' component with $t_{1}$
on the positive time-branch while $t_{2}$ on the negative time-branch
is denoted as $G_{\mu\nu}^{<}(x_{1},x_{2})$ meaning that $t_{2}$
is always later than $t_{1}$ on the CTP contour as shown in Fig.
\ref{fig:ctp}(b). Analogously, $G_{\mu\nu}^{>}(x_{1},x_{2})$ denotes
the '$-+$' component with $t_{1}$ on the negative time-branch and
$t_{2}$ on the positive time-branch as shown in Fig. \ref{fig:ctp}(c),
while $G_{\mu\nu}^{\overline{F}}(x_{1},x_{2})$ denotes the '$--$'
component with both $t_{1}$ and $t_{2}$ on the negative time-branch
as shown in Fig. \ref{fig:ctp}(d).

The Wigner function can be defined from $G_{\mu\nu}^{<}(x_{1},x_{2})$
by taking a Fourier transform with respect to the relative position
$y=x_{1}-x_{2}$,
\begin{eqnarray}
G_{\mu\nu}^{<}(x,p) & \equiv & \int d^{4}y\,e^{ip\cdot y/\hbar}G_{\mu\nu}^{<}(x_{1},x_{2})\nonumber \\
 & = & \int d^{4}y\,e^{ip\cdot y/\hbar}\left\langle A_{\nu}^{\dagger}(x_{2})A_{\mu}(x_{1})\right\rangle .\label{eq:definition_G<}
\end{eqnarray}
Inserting the quantized field (\ref{eq:a-quantization}) into the
definition of the Wigner function (\ref{eq:definition_G<}), we obtain
\begin{eqnarray}
G_{\mu\nu}^{<}(x,p) & = & 2\pi\hbar\sum_{\lambda_{1},\lambda_{2}}\delta\left(p^{2}-m_{V}^{2}\right)\nonumber \\
 &  & \times\left\{ \theta(p^{0})\epsilon_{\mu}\left(\lambda_{1},{\bf p}\right)\epsilon_{\nu}^{\ast}\left(\lambda_{2},{\bf p}\right)f_{\lambda_{1}\lambda_{2}}(x,{\bf p})\right.\nonumber \\
 &  & +\theta(-p^{0})\epsilon_{\mu}^{\ast}\left(\lambda_{1},-{\bf p}\right)\epsilon_{\nu}\left(\lambda_{2},-{\bf p}\right)\nonumber \\
 &  & \left.\times\left[\delta_{\lambda_{2}\lambda_{1}}+f_{\lambda_{2}\lambda_{1}}(x,-{\bf p})\right]\right\} ,\label{eq:wigner-func-vm}
\end{eqnarray}
where the gradient expansion has been taken with spatial gradient
terms being dropped, and the MVSD for the vector meson is defined
as 
\begin{eqnarray}
f_{\lambda_{1}\lambda_{2}}(x,{\bf p}) & \equiv & \int\frac{d^{4}u}{2(2\pi\hbar)^{3}}\delta(p\cdot u)e^{-iu\cdot x/\hbar}\nonumber \\
 &  & \times\left\langle a_{V}^{\dagger}\left(\lambda_{2},{\bf p}-\frac{{\bf u}}{2}\right)a_{V}\left(\lambda_{1},{\bf p}+\frac{{\bf u}}{2}\right)\right\rangle .
\end{eqnarray}
One can check that $f_{\lambda_{1}\lambda_{2}}(x,{\bf p})$ is a Hermitian
matrix, i.e. $f_{\lambda_{1}\lambda_{2}}^{*}(x,{\bf p})=f_{\lambda_{2}\lambda_{1}}(x,{\bf p})$.
Similarly we can define another Wigner function from $G_{\mu\nu}^{>}(x_{1},x_{2})$
\begin{eqnarray}
G_{\mu\nu}^{>}(x,p) & \equiv & \int d^{4}y\,e^{ip\cdot y/\hbar}\left\langle A_{\mu}(x_{1})A_{\nu}^{\dagger}(x_{2})\right\rangle .\nonumber \\
 & = & 2\pi\hbar\sum_{\lambda_{1},\lambda_{2}}\delta\left(p^{2}-m_{V}^{2}\right)\nonumber \\
 &  & \times\left\{ \theta(p^{0})\epsilon_{\mu}\left(\lambda_{1},{\bf p}\right)\epsilon_{\nu}^{\ast}\left(\lambda_{2},{\bf p}\right)\left[\delta_{\lambda_{1}\lambda_{2}}+f_{\lambda_{1}\lambda_{2}}(x,{\bf p})\right]\right.\nonumber \\
 &  & \left.+\theta(-p^{0})\epsilon_{\mu}^{\ast}\left(\lambda_{1},-{\bf p}\right)\epsilon_{\nu}\left(\lambda_{2},-{\bf p}\right)f_{\lambda_{2}\lambda_{1}}(x,-{\bf p})\right\} .\label{eq:wigner-func-vm1}
\end{eqnarray}
Note that $G_{\mu\nu}^{>}(x,p)$ can be obtained by replacing $f_{\lambda_{1}\lambda_{2}}\rightarrow\delta_{\lambda_{1}\lambda_{2}}+f_{\lambda_{1}\lambda_{2}}$
and $\delta_{\lambda_{2}\lambda_{1}}+f_{\lambda_{2}\lambda_{1}}\rightarrow f_{\lambda_{2}\lambda_{1}}$
from $G_{\mu\nu}^{<}(x,p)$.

\section{Kadanoff-Baym equations for vector mesons}

\label{sec:kb-vector}The Wigner functions for massless vector particles
such as gluons and photons \citep{Elze:1986hq,Blaizot:2001nr,Wang:2001dm,Huang:2020kik,Hattori:2020gqh,Muller:2021hpe}
have been studies for many years, but to our knowledge there are few
works about Wigner functions for massive vector mesons in the context
of spin polarization (see Ref. \citep{Weickgenannt:2022jes} for a
recent one). In this section, we will derive the Boltzmann equation
for vector mesons' Wigner functions from two-point Green functions
on the CTP. The starting point is the KB equations 
\begin{eqnarray}
 &  & L_{\eta}^{\mu}(x_{1})G^{<,\eta\nu}(x_{1},x_{2})\nonumber \\
 & = & -\frac{i\hbar}{2}\int d^{4}x^{\prime}\left[\Sigma_{\ \ \ \ \alpha}^{<,\mu}(x_{1},x^{\prime})G^{>,\alpha\nu}(x^{\prime},x_{2})-\Sigma_{\ \ \ \ \alpha}^{>,\mu}(x_{1},x^{\prime})G^{<,\alpha\nu}(x^{\prime},x_{2})\right],\label{eq:kbe-G<V}
\end{eqnarray}
and 
\begin{eqnarray}
 &  & L_{\eta}^{\nu}(x_{2})G^{<,\mu\eta}(x_{1},x_{2})\nonumber \\
 & = & -\frac{i\hbar}{2}\int d^{4}x^{\prime}\left[G_{\ \ \ \ \alpha}^{<,\mu}(x_{1},x^{\prime})\Sigma^{>,\alpha\nu}(x^{\prime},x_{2})-G_{\ \ \ \ \alpha}^{>,\mu}(x_{1},x^{\prime})\Sigma^{<,\alpha\nu}(x^{\prime},x_{2})\right].\label{eq:kbe-G<V-1}
\end{eqnarray}
Equations (\ref{eq:kbe-G<V}) and (\ref{eq:kbe-G<V-1}) are the result
of the quasiparticle approximation \citep{Sheng:2021kfc}. Note that
the integrations over $x$ in Eqs. (\ref{eq:kbe-G<V}) and (\ref{eq:kbe-G<V-1})
are ordinary ones.

\begin{figure}[h]
\includegraphics[scale=0.8]{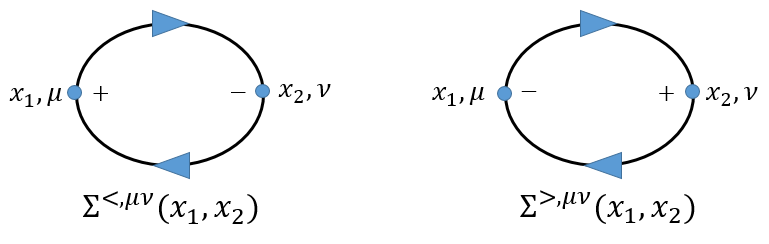}

\caption{\label{fig:self-energy-vm}The self-energies $\Sigma^{<,\mu\nu}$
and $\Sigma^{>,\mu\nu}$ of vector mesons from quark loops in the
quark-meson model. Two quark propagators in the loop may have different
flavors corresponding to the vector meson that is not flavor neutral.}
\end{figure}

We consider the coupling between the vector meson and the quark-antiquark
in the quark-meson model \citep{Manohar:1983md,Fernandez:1993hx,Li:1997gd,Zhao:1998fn,Zacchi:2015lwa,Zacchi:2016tjw}.
Then at lowest order in the coupling constant, the self-energies are
from quark loops as shown in Fig. \ref{fig:self-energy-vm}
\begin{eqnarray}
\Sigma^{<,\mu\nu}(x_{1},x_{2}) & = & -\text{Tr}\left[i\Gamma^{\mu}S^{<}(x_{1},x_{2})i\Gamma^{\nu}S^{>}(x_{2},x_{1})\right],\nonumber \\
\Sigma^{>,\mu\nu}(x_{1},x_{2}) & = & -\text{Tr}\left[i\Gamma^{\mu}S^{>}(x_{1},x_{2})i\Gamma^{\nu}S^{<}(x_{2},x_{1})\right],\label{eq:self-en}
\end{eqnarray}
where $S^{>}(x_{1},x_{2})$ and $S^{<}(x_{1},x_{2})$ are two-point
Green functions of quarks, $i\Gamma^{\mu}$ denotes the vertex of
the vector meson and quark-antiquark, and the overall minus signs
arise from quark loops. Inserting the self-energies (\ref{eq:self-en})
into Eq. (\ref{eq:kbe-G<V}) and taking a Fourier transform with respect
to the difference $y=x_{1}-x_{2}$, we obtain the KB equation for
the Wigner function as 
\begin{eqnarray}
 &  & \left\{ g_{\eta}^{\mu}\left[-\left(p^{2}-m_{V}^{2}-\frac{\hbar^{2}}{4}\partial_{x}^{2}\right)-i\hbar p\cdot\partial_{x}\right]\right.\nonumber \\
 &  & \left.-\frac{\hbar^{2}}{4}\partial_{x}^{\mu}\partial_{\eta}^{x}+p^{\mu}p_{\eta}+\frac{1}{2}i\hbar\left(p_{\eta}\partial_{x}^{\mu}+p^{\mu}\partial_{\eta}^{x}\right)\right\} G^{<,\eta\nu}(x,p)\nonumber \\
 & = & -\frac{i\hbar}{2}\int\frac{d^{4}p^{\prime}}{(2\pi\hbar)^{4}}\left\{ \text{Tr}\left[\Gamma^{\mu}S^{<}\left(x,p+p^{\prime}\right)\Gamma_{\alpha}S^{>}\left(x,p^{\prime}\right)\right]G^{>,\alpha\nu}\left(x,p\right)\right.\nonumber \\
 &  & \left.-\text{Tr}\left[\Gamma^{\mu}S^{>}\left(x,p+p^{\prime}\right)\Gamma_{\alpha}S^{<}\left(x,p^{\prime}\right)\right]G^{<,\alpha\nu}\left(x,p\right)\right\} \nonumber \\
 &  & -\frac{\hbar^{2}}{4}\int\frac{d^{4}p^{\prime}}{(2\pi\hbar)^{4}}\left[\left\{ \text{Tr}\left[\Gamma^{\mu}S^{<}\left(x,p+p^{\prime}\right)\Gamma_{\alpha}S^{>}\left(x,p^{\prime}\right)\right],G^{>,\alpha\nu}\left(x,p\right)\right\} _{\text{P.B.}}\right.\nonumber \\
 &  & \left.-\left\{ \text{Tr}\left[\Gamma^{\mu}S^{>}\left(x,p+p^{\prime}\right)\Gamma_{\alpha}S^{<}\left(x,p^{\prime}\right)\right],G^{<,\alpha\nu}\left(x,p\right)\right\} _{\text{P.B.}}\right],\label{eq:kbe-right}
\end{eqnarray}
where the Poisson bracket involves space-time and momentum gradients
and is defined as 
\begin{equation}
\left\{ A,B\right\} _{\text{P.B.}}\equiv(\partial_{x}^{\mu}A)(\partial_{\mu}^{p}B)-(\partial_{p}^{\mu}A)(\partial_{\mu}^{x}B).
\end{equation}
In the same way, we obtain from Eq. (\ref{eq:kbe-G<V-1}) another
KB equation for the Wigner function 
\begin{eqnarray}
 &  & \left\{ g_{\eta}^{\nu}\left[-\left(p^{2}-m_{V}^{2}-\frac{\hbar^{2}}{4}\partial_{x}^{2}\right)+i\hbar p\cdot\partial_{x}\right]\right.\nonumber \\
 &  & \left.-\frac{\hbar^{2}}{4}\partial_{x}^{\nu}\partial_{\eta}^{x}+p^{\nu}p_{\eta}-\frac{1}{2}i\hbar\left(p_{\eta}\partial_{x}^{\nu}+p^{\nu}\partial_{\eta}^{x}\right)\right\} G^{<,\mu\eta}(x,p)\nonumber \\
 & = & -\frac{i\hbar}{2}\int\frac{d^{4}p^{\prime}}{(2\pi\hbar)^{4}}\left\{ G_{\ \ \ \ \alpha}^{<,\mu}(x,p)\text{Tr}\left[\Gamma^{\alpha}S^{>}\left(x,p+p^{\prime}\right)\Gamma^{\nu}S^{<}\left(x,p^{\prime}\right)\right]\right.\nonumber \\
 &  & \left.-G_{\ \ \ \ \alpha}^{>,\mu}\left(x,p\right)\text{Tr}\left[\Gamma^{\alpha}S^{<}\left(x,p+p^{\prime}\right)\Gamma^{\nu}S^{>}\left(x,p^{\prime}\right)\right]\right\} \nonumber \\
 &  & -\frac{\hbar^{2}}{4}\int\frac{d^{4}p^{\prime}}{(2\pi\hbar)^{4}}\left[\left\{ G_{\ \ \ \ \alpha}^{<,\mu}(x,p),\text{Tr}\left[\Gamma^{\alpha}S^{>}\left(x,p+p^{\prime}\right)\Gamma^{\nu}S^{<}\left(x,p^{\prime}\right)\right]\right\} _{\text{P.B.}}\right.\nonumber \\
 &  & \left.-\left\{ G_{\ \ \ \ \alpha}^{>,\mu}\left(x,p\right),\text{Tr}\left[\Gamma^{\alpha}S^{<}\left(x,p+p^{\prime}\right)\Gamma^{\nu}S^{>}\left(x,p^{\prime}\right)\right]\right\} _{\text{P.B.}}\right].\label{eq:kbe-left}
\end{eqnarray}
Taking the difference between Eq. (\ref{eq:kbe-right}) and (\ref{eq:kbe-left}),
we are able to derive the Boltzmann equation for the Wigner function
at the leading order 
\begin{eqnarray}
 &  & p\cdot\partial_{x}G^{<,\mu\nu}(x,p)-\frac{1}{4}\left[p^{\mu}\partial_{\eta}^{x}G^{<,\eta\nu}(x,p)+p^{\nu}\partial_{\eta}^{x}G^{<,\mu\eta}(x,p)\right]\nonumber \\
 & = & \frac{1}{4}\int\frac{d^{4}p^{\prime}}{(2\pi\hbar)^{4}}\left\{ \text{Tr}\left[\Gamma^{\mu}S^{<}\left(x,p+p^{\prime}\right)\Gamma_{\alpha}S^{>}\left(x,p^{\prime}\right)\right]G^{>,\alpha\nu}\left(x,p\right)\right.\nonumber \\
 &  & \left.-\text{Tr}\left[\Gamma^{\mu}S^{>}\left(x,p+p^{\prime}\right)\Gamma_{\alpha}S^{<}\left(x,p^{\prime}\right)\right]G^{<,\alpha\nu}\left(x,p\right)\right\} \nonumber \\
 &  & +\frac{1}{4}\int\frac{d^{4}p^{\prime}}{(2\pi\hbar)^{4}}\left\{ G_{\ \ \ \ \alpha}^{>,\mu}\left(x,p\right)\text{Tr}\left[\Gamma^{\alpha}S^{<}\left(x,p+p^{\prime}\right)\Gamma^{\nu}S^{>}\left(x,p^{\prime}\right)\right]\right.\nonumber \\
 &  & \left.-G_{\ \ \ \ \alpha}^{<,\mu}(x,p)\text{Tr}\left[\Gamma^{\alpha}S^{>}\left(x,p+p^{\prime}\right)\Gamma^{\nu}S^{<}\left(x,p^{\prime}\right)\right]\right\} ,\label{eq:boltzmann-eq}
\end{eqnarray}
where we have neglected terms with Poisson brackets and those proportional
to $p_{\eta}$ in the left-hand-side since their contraction with
the leading-order $G^{<,\eta\nu}(x,p)$ and $G^{<,\mu\eta}(x,p)$
is vanishing. In the next section we will rewrite the above Boltzmann
equation in terms of MVSDs for vector mesons, quarks and antiquarks.

\section{Spin density matrix for quark coalescence and dissociation}

\label{sec:spin-density-matrix}The two-point Green functions $S^{>}\left(x,p\right)$
and $S^{<}\left(x,p\right)$ for quarks are given by \citep{Sheng:2021kfc}
\begin{eqnarray}
S^{<}(x,p) & = & -(2\pi\hbar)\theta(p_{0})\delta(p^{2}-m_{q}^{2})\sum_{r,s}u(r,\mathbf{p})\overline{u}(s,\mathbf{p})f_{rs}^{(+)}(x,\mathbf{p})\nonumber \\
 &  & -(2\pi\hbar)\theta(-p_{0})\delta(p^{2}-m_{q}^{2})\sum_{r,s}v(s,-\mathbf{p})\overline{v}(r,-\mathbf{p})\left[\delta_{rs}-f_{rs}^{(-)}(x,-\mathbf{p})\right],\nonumber \\
S^{>}(x,p) & = & (2\pi\hbar)\theta(p_{0})\delta(p^{2}-m_{q}^{2})\sum_{r,s}u(r,\mathbf{p})\overline{u}(s,\mathbf{p})\left[\delta_{rs}-f_{rs}^{(+)}(x,\mathbf{p})\right]\nonumber \\
 &  & +(2\pi\hbar)\theta(-p_{0})\delta(p^{2}-m_{q}^{2})\sum_{r,s}v(s,-\mathbf{p})\overline{v}(r,-\mathbf{p})f_{rs}^{(-)}(x,-\mathbf{p}),\label{eq:G_F}
\end{eqnarray}
where $f_{rs}^{(+)}$ and $f_{rs}^{(-)}$ are MVSD for quarks and
antiquarks respectively. We can parameterize them as 
\begin{eqnarray}
f_{rs}^{(+)}(x,\mathbf{p}) & = & \frac{1}{2}f_{q}(x,\mathbf{p})\left[\delta_{rs}-P_{\mu}^{q}(x,\mathbf{p})n_{j}^{(+)\mu}(\mathbf{p})\tau_{rs}^{j}\right],\nonumber \\
f_{rs}^{(-)}(x,-\mathbf{p}) & = & \frac{1}{2}f_{\overline{q}}(x,-\mathbf{p})\left[\delta_{rs}-P_{\mu}^{\overline{q}}(x,-\mathbf{p})n_{j}^{(-)\mu}(\mathbf{p})\tau_{rs}^{j}\right],\label{eq:f-rs-pol}
\end{eqnarray}
where $f_{q}(x,\mathbf{p})$ and $f_{\overline{q}}(x,-\mathbf{p})$
are MVSDs for quarks and antiquarks respectively, and $P_{q}^{\mu}(x,\mathbf{p})$
and $P_{\overline{q}}^{\mu}(x,-\mathbf{p})$ are polarization four-vectors
for quarks and antiquarks respectively. The spin direction four-vectors
for quarks and antiquarks are given by 
\begin{eqnarray}
n_{j}^{(+)\mu}(\mathbf{p}) & \equiv & n^{\mu}(\mathbf{n}_{j},\mathbf{p},m_{q})=\left(\frac{\mathbf{n}_{j}\cdot{\bf p}}{m_{q}},\mathbf{n}_{j}+\frac{(\mathbf{n}_{j}\cdot{\bf p}){\bf p}}{m_{q}(E_{{\bf p}}^{q}+m_{q})}\right),\nonumber \\
n_{j}^{(-)\mu}(\mathbf{p}) & \equiv & n^{\mu}(\mathbf{n}_{j},-\mathbf{p},m_{\overline{q}})=\left(-\frac{\mathbf{n}_{j}\cdot{\bf p}}{m_{\overline{q}}},\mathbf{n}_{j}+\frac{(\mathbf{n}_{j}\cdot{\bf p}){\bf p}}{m_{\overline{q}}(E_{{\bf p}}^{\overline{q}}+m_{\overline{q}})}\right),
\end{eqnarray}
where $\mathbf{n}_{j}$ for $j=1,2,3$ are three basis unit vectors
that form a Cartesian coordinate system in the particle's rest frame
with $\mathbf{n}_{3}$ being the spin quantization direction, and
$n_{j}^{(+)\mu}$ and $n_{j}^{(-)\mu}$ are the Lorentz transformed
four-vectors of $\mathbf{n}_{j}$ for quarks and antiquarks respectively
which obey the sum rules 
\begin{eqnarray}
n_{j}^{(+)\mu}(\mathbf{p})n_{j}^{(+)\nu}(\mathbf{p}) & = & -\left(g^{\mu\nu}-\frac{p^{\mu}p^{\nu}}{m_{q}^{2}}\right),\nonumber \\
n_{j}^{(-)\mu}(\mathbf{p})n_{j}^{(-)\nu}(\mathbf{p}) & = & -\left(g^{\mu\nu}-\frac{\overline{p}^{\mu}\overline{p}^{\nu}}{m_{\overline{q}}^{2}}\right),
\end{eqnarray}
where $p^{\mu}=(E_{p}^{q},\mathbf{p})$ and $\overline{p}^{\mu}=(E_{p}^{\overline{q}},-\mathbf{p})$.
We note that $f_{rs}^{(+)}(x,\mathbf{p})$ and $f_{rs}^{(-)}(x,-\mathbf{p})$
are actually the transpose of those MVSDs defined in Eqs. (117)-(118)
of Ref. \citep{Sheng:2021kfc} in spin indices. We can flip the sign
of the three-momentum, $\mathbf{p}\rightarrow-\mathbf{p}$, in $f_{rs}^{(-)}(x,-\mathbf{p})$
to obtain 
\begin{equation}
f_{rs}^{(-)}(x,\mathbf{p})=\frac{1}{2}f_{\overline{q}}(x,\mathbf{p})\left[\delta_{rs}-P_{\mu}^{\overline{q}}(x,\mathbf{p})n_{j}^{(-)\mu}(-\mathbf{p})\tau_{rs}^{j}\right],
\end{equation}
where $n_{j}^{(-)\mu}(-\mathbf{p})$ has the same form as $n_{j}^{(+)\mu}(\mathbf{p})$
except the quark mass. Note that in the self-energy (\ref{eq:self-en})
of the vector meson that is not flavor neutral, $S^{<}(x,p)$ and
$S^{>}(x,p)$ may involve different flavors of quarks and antiquarks.

Inserting $S^{<}(x,p)$, $S^{>}(x,p)$, $G^{<,\mu\nu}(x,p)$, and
$G^{>,\mu\nu}(x,p)$ in Eqs. (\ref{eq:G_F}), (\ref{eq:wigner-func-vm})
and (\ref{eq:wigner-func-vm1}) into Eq. (\ref{eq:boltzmann-eq}),
the Boltzmann equation can be put into the following form 
\begin{eqnarray}
 &  & p\cdot\partial_{x}G^{<,\mu\nu}(x,p)-\frac{1}{4}\left[p^{\mu}\partial_{\eta}^{x}G^{<,\eta\nu}(x,p)+p^{\nu}\partial_{\eta}^{x}G^{<,\mu\eta}(x,p)\right]\nonumber \\
 & = & \frac{1}{4(2\pi\hbar)}\int d^{4}p^{\prime}\delta(p^{\prime2}-m_{\overline{q}}^{2})\delta\left[(p+p^{\prime})^{2}-m_{q}^{2}\right]\delta(p^{2}-m_{V}^{2})\nonumber \\
 &  & \times\left\{ \theta(p_{0}^{\prime})\theta\left(p_{0}+p_{0}^{\prime}\right)\theta(p_{0})I_{+++}\right.\nonumber \\
 &  & +\theta(p_{0}^{\prime})\theta\left(p_{0}+p_{0}^{\prime}\right)\theta(-p_{0})I_{++-}\nonumber \\
 &  & +\theta(p_{0}^{\prime})\theta\left(-p_{0}-p_{0}^{\prime}\right)\theta(-p_{0})I_{+--}\nonumber \\
 &  & +\theta(-p_{0}^{\prime})\theta\left(p_{0}+p_{0}^{\prime}\right)\theta(p_{0})I_{-++}\nonumber \\
 &  & +\theta(-p_{0}^{\prime})\theta\left(-p_{0}-p_{0}^{\prime}\right)\theta(p_{0})I_{--+}\nonumber \\
 &  & \left.+\theta(-p_{0}^{\prime})\theta\left(-p_{0}-p_{0}^{\prime}\right)\theta(-p_{0})I_{---}\right\} .\label{eq:boltzmann-all}
\end{eqnarray}
The terms $I_{ijk}$, with $i,j,k=\pm$ representing the positive/negative
energy, correspond to all possible processes at lowest order in the
coupling constant, as shown in Table \ref{tab:collision-terms}. In
Eq. (\ref{eq:boltzmann-all}), $I_{-+-}$ and $I_{+-+}$ are absent
due to incompatibility of theta functions, and $I_{-++}$ and $I_{+--}$
contain the coalescence of quark-antiquark to the vector meson and
vice versa, but $I_{-++}$ corresponds to the positive energy sector
of the two-point function for the vector meson while $I_{+--}$ corresponds
to the negative energy sector. All terms except $I_{-++}$ and $I_{+--}$
are vanishing for on-shell quarks, antiquarks and mesons at the one-loop
level of the selfenergy. We distinguish $m_{q}$ from $m_{\overline{q}}$
in Eq. (\ref{eq:boltzmann-all}) since the quark and antiquark may
have different flavors for the vector meson that is not flavor neutral
so the meson and its antiparticle are not the same particle.

\begin{table}
\begin{tabular}{|c|c|c|c|c|c|}
\hline 
$I_{+++}$ & $I_{++-}$ & $I_{+--}$ & $I_{-++}$ & $I_{--+}$ & $I_{---}$\tabularnewline
\hline 
$q\rightarrow q+M$ & $q+\overline{M}\rightarrow q$ & $\overline{M}\rightarrow q+\overline{q}$ & $q+\overline{q}\rightarrow M$ & $\overline{q}\rightarrow\overline{q}+M$ & $\overline{q}+\overline{M}\rightarrow\overline{q}$\tabularnewline
$q+M\rightarrow q$ & $q\rightarrow q+\overline{M}$ & $q+\overline{q}\rightarrow\overline{M}$ & $M\rightarrow q+\overline{q}$ & $\overline{q}+M\rightarrow\overline{q}$ & $\overline{q}\rightarrow\overline{q}+\overline{M}$\tabularnewline
\hline 
\end{tabular}

\caption{\label{tab:collision-terms}Collision terms in the Boltzmann equation.
All terms except $I_{-++}$ and $I_{+--}$ are vanishing for on-shell
quarks, antiquarks and mesons at the one-loop level of the selfenergy.}
\end{table}

In this paper, we are interested in the contribution from the coalescence
and dissociation processes corresponding to $I_{-++}$. The coalescence
is regarded as one of the main processes for particle production in
heavy-ion collisions \citep{Greco:2003xt,Fries:2003vb,Greco:2003mm,Fries:2003kq,Greco:2003vf,Zhao:2020wcd}.
So the spin Boltzmann equation for the vector meson's MVSD reads 
\begin{eqnarray}
 &  & p\cdot\partial_{x}f_{\lambda_{1}\lambda_{2}}(x,\mathbf{p})\nonumber \\
 & = & \frac{1}{16}\sum_{r_{1},s_{1},r_{2},s_{2},\lambda_{1}^{\prime},\lambda_{2}^{\prime}}\int\frac{d^{3}\mathbf{p}^{\prime}}{(2\pi\hbar)^{3}}\frac{1}{E_{p^{\prime}}^{\overline{q}}E_{{\bf p}-{\bf p}^{\prime}}^{q}}2\pi\hbar\delta\left(E_{p}^{V}-E_{p^{\prime}}^{\overline{q}}-E_{{\bf p}-{\bf p}^{\prime}}^{q}\right)\nonumber \\
 &  & \times\left\{ \delta_{\lambda_{2}\lambda_{2}^{\prime}}\epsilon_{\mu}^{\ast}(\lambda_{1},{\bf p})\epsilon^{\alpha}\left(\lambda_{1}^{\prime},{\bf p}\right)\text{Tr}\left[\Gamma_{\alpha}v(s_{1},{\bf p}^{\prime})\overline{v}(r_{1},{\bf p}^{\prime})\Gamma^{\mu}u(r_{2},\mathbf{p}-{\bf p}^{\prime})\overline{u}(s_{2},\mathbf{p}-{\bf p}^{\prime})\right]\right.\nonumber \\
 &  & +\left.\delta_{\lambda_{1}\lambda_{1}^{\prime}}\epsilon_{\nu}(\lambda_{2},{\bf p})\epsilon_{\alpha}^{\ast}\left(\lambda_{2}^{\prime},{\bf p}\right)\text{Tr}\left[\Gamma^{\nu}v(s_{1},{\bf p}^{\prime})\overline{v}(r_{1},{\bf p}^{\prime})\Gamma^{\alpha}u(r_{2},\mathbf{p}-{\bf p}^{\prime})\overline{u}(s_{2},\mathbf{p}-{\bf p}^{\prime})\right]\right\} \nonumber \\
 &  & \times\left\{ f_{r_{1}s_{1}}^{(-)}(x,{\bf p}^{\prime})f_{r_{2}s_{2}}^{(+)}(x,\mathbf{p}-\mathbf{p}^{\prime})\left[\delta_{\lambda_{1}^{\prime}\lambda_{2}^{\prime}}+f_{\lambda_{1}^{\prime}\lambda_{2}^{\prime}}(x,\mathbf{p})\right]\right.\nonumber \\
 &  & \left.-\left[\delta_{r_{1}s_{1}}-f_{r_{1}s_{1}}^{(-)}(x,{\bf p}^{\prime})\right]\left[\delta_{r_{2}s_{2}}-f_{r_{2}s_{2}}^{(+)}(x,\mathbf{p}-\mathbf{p}^{\prime})\right]f_{\lambda_{1}^{\prime}\lambda_{2}^{\prime}}(x,\mathbf{p})\right\} ,\label{eq:boltz-mvsd-1}
\end{eqnarray}
where $\lambda_{1}$, $\lambda_{2}$, $\lambda_{1}^{\prime}$ and
$\lambda_{2}^{\prime}$ denote the spin states along the spin quantization
direction, and $\Gamma^{\alpha}$ is the $q\overline{q}V$ vertex
given by 
\begin{equation}
\Gamma^{\alpha}\approx g_{V}B(\mathbf{p}-\mathbf{p}^{\prime},\mathbf{p}^{\prime})\gamma^{\alpha},\label{eq:qq-bar-vertex}
\end{equation}
where $g_{V}$ is the coupling constant of the vector meson and quark-antiquark,
and $B(\mathbf{p}-\mathbf{p}^{\prime},\mathbf{p}^{\prime})$ denotes
the Bethe-Salpeter wave function of the $\phi$ meson \citep{Xu:2019ilh,Xu:2021mju}
in the following parametrization form 
\begin{equation}
B(\mathbf{p}-\mathbf{p}^{\prime},\mathbf{p}^{\prime})=\frac{1-\exp\left\{ -\left[(E_{\mathbf{p}-\mathbf{p}^{\prime}}^{s}-E_{\mathbf{p}^{\prime}}^{\overline{s}})^{2}-(\mathbf{p}-2\mathbf{p}^{\prime})^{2}\right]/\sigma^{2}\right\} }{\left[(E_{\mathbf{p}-\mathbf{p}^{\prime}}^{s}-E_{\mathbf{p}^{\prime}}^{\overline{s}})^{2}-(\mathbf{p}-2\mathbf{p}^{\prime})^{2}\right]/\sigma^{2}},
\end{equation}
with $\sigma\approx0.522$ GeV being the width parameter of the wave
function. The derivation of Eq. (\ref{eq:boltz-mvsd-1}) is given
in Appendix \ref{sec:coalescence}. We see that there is a gain term
and a loss term in Eq. (\ref{eq:boltz-mvsd-1}). In heavy ion collisions,
the distribution functions are normally much less than 1, $f_{\lambda_{1}\lambda_{2}}(x,\mathbf{p})\sim f_{rs}^{(+)}\sim f_{rs}^{(-)}\ll1$,
so Eq. (\ref{eq:boltz-mvsd-1}) can be approximated as 
\begin{equation}
\frac{p}{E_{p}^{V}}\cdot\partial_{x}f_{\lambda_{1}\lambda_{2}}(x,\mathbf{p})\approx R_{\lambda_{1}\lambda_{2}}^{\mathrm{coal}}(\mathbf{p})-R^{\mathrm{diss}}(\mathbf{p})f_{\lambda_{1}\lambda_{2}}(x,\mathbf{p}),\label{eq:boltzmann-v}
\end{equation}
where $R_{\lambda_{1}\lambda_{2}}^{\mathrm{coal}}$ and $R_{\lambda_{1}\lambda_{2}}^{\mathrm{diss}}$
denote the coalescence and dissociation rates for the vector meson,
i.e. the rates of $q+\overline{q}\rightarrow M$ and $M\rightarrow q+\overline{q}$
respectively, defined as 
\begin{eqnarray}
R_{\lambda_{1}\lambda_{2}}^{\mathrm{coal}}(\mathbf{p}) & = & \frac{1}{8(2\pi\hbar)^{2}}\sum_{r_{1},s_{1},r_{2},s_{2}}\int d^{3}\mathbf{p}^{\prime}\frac{1}{E_{p^{\prime}}^{\overline{q}}E_{{\bf p}-{\bf p}^{\prime}}^{q}E_{p}^{V}}\nonumber \\
 &  & \times\delta\left(E_{p}^{V}-E_{p^{\prime}}^{\overline{q}}-E_{{\bf p}-{\bf p}^{\prime}}^{q}\right)\epsilon_{\alpha}^{\ast}(\lambda_{1},{\bf p})\epsilon_{\beta}(\lambda_{2},{\bf p})\nonumber \\
 &  & \times\text{Tr}\left[\Gamma^{\beta}v(s_{1},{\bf p}^{\prime})\overline{v}(r_{1},{\bf p}^{\prime})\Gamma^{\alpha}u(r_{2},\mathbf{p}-{\bf p}^{\prime})\overline{u}(s_{2},\mathbf{p}-{\bf p}^{\prime})\right]\nonumber \\
 &  & \times f_{r_{1}s_{1}}^{(-)}(x,{\bf p}^{\prime})f_{r_{2}s_{2}}^{(+)}(x,\mathbf{p}-{\bf p}^{\prime}),\label{eq:coal}\\
R^{\mathrm{diss}}(\mathbf{p}) & = & -\frac{1}{12(2\pi\hbar)^{2}}\sum_{r_{1},r_{2}}\int d^{3}\mathbf{p}^{\prime}\frac{1}{E_{p^{\prime}}^{\overline{q}}E_{{\bf p}-{\bf p}^{\prime}}^{q}E_{p}^{V}}\nonumber \\
 &  & \times\delta\left(E_{p}^{V}-E_{p^{\prime}}^{\overline{q}}-E_{{\bf p}-{\bf p}^{\prime}}^{q}\right)\left(g_{\alpha\beta}-\frac{p_{\alpha}p_{\beta}}{m_{V}^{2}}\right)\nonumber \\
 &  & \times\text{Tr}\left\{ \Gamma^{\beta}\left(p^{\prime}\cdot\gamma-m_{\overline{q}}\right)\Gamma^{\alpha}\left[(p-p^{\prime})\cdot\gamma+m_{q}\right]\right\} .\label{eq:dis-loss}
\end{eqnarray}
Note that $R^{\mathrm{diss}}(\mathbf{p})$ does not depend on the
MVSDs of the quark, antiquark and the vector meson, therefore it is
independent of the quark polarization. Schematically the formal solution
to Eq. (\ref{eq:boltzmann-v}) reads 
\begin{eqnarray}
f_{\lambda_{1}\lambda_{2}}(x,\mathbf{p}) & \sim & \frac{R_{\lambda_{1}\lambda_{2}}^{\mathrm{coal}}(\mathbf{p})}{R^{\mathrm{diss}}(\mathbf{p})}\left[1-\exp\left(-R^{\mathrm{diss}}(\mathbf{p})\Delta t\right)\right]\nonumber \\
 & \sim & \begin{cases}
R_{\lambda_{1}\lambda_{2}}^{\mathrm{coal}}(\mathbf{p})\Delta t, & \mathrm{for}\;\Delta t\ll1/R^{\mathrm{diss}}(\mathbf{p})\\
\frac{R_{\lambda_{1}\lambda_{2}}^{\mathrm{coal}}(\mathbf{p})}{R^{\mathrm{diss}}(\mathbf{p})}, & \mathrm{for}\;\Delta t\gg1/R^{\mathrm{diss}}(\mathbf{p})
\end{cases}
\end{eqnarray}
if $f_{\lambda_{1}\lambda_{2}}(x,\mathbf{p})$ for the vector meson
at the initial time is assumed to be zero, where $\Delta t$ is the
formation time of the vector meson.

The spin density matrix element $\rho_{\lambda_{1}\lambda_{2}}^{V}$
is assumed to be proportional to $f_{\lambda_{1}\lambda_{2}}(x,\mathbf{p})$
which is $R_{\lambda_{1}\lambda_{2}}^{\mathrm{coal}}(\mathbf{p})\Delta t$
if $\Delta t\ll1/R^{\mathrm{diss}}(\mathbf{p})$ or $R_{\lambda_{1}\lambda_{2}}^{\mathrm{coal}}(\mathbf{p})/R^{\mathrm{diss}}(\mathbf{p})$
if $\Delta t\gg1/R^{\mathrm{diss}}(\mathbf{p})$. In both cases, $\rho_{\lambda_{1}\lambda_{2}}^{V}$
is proportional to $R_{\lambda_{1}\lambda_{2}}^{\mathrm{coal}}(\mathbf{p})$
times a constant independent of the spin states of the vector meson.
Here we assume that the coalescence of the vector meson takes place
in a relatively short time, so we have 
\begin{eqnarray}
\rho_{\lambda_{1}\lambda_{2}}^{V}(x,{\bf p}) & \approx & \frac{\Delta t}{8}\sum_{r_{1},s_{1},r_{2},s_{2}}\int\frac{d^{3}\mathbf{p}^{\prime}}{(2\pi\hbar)^{3}}\frac{1}{E_{p^{\prime}}^{\overline{q}}E_{{\bf p}-{\bf p}^{\prime}}^{q}E_{p}^{V}}\nonumber \\
 &  & \times2\pi\hbar\delta\left(E_{p}^{V}-E_{p^{\prime}}^{\overline{q}}-E_{{\bf p}-{\bf p}^{\prime}}^{q}\right)\epsilon_{\alpha}^{\ast}(\lambda_{1},{\bf p})\epsilon_{\beta}(\lambda_{2},{\bf p})\nonumber \\
 &  & \times\text{Tr}\left[\Gamma^{\beta}v(s_{1},{\bf p}^{\prime})\overline{v}(r_{1},{\bf p}^{\prime})\Gamma^{\alpha}u(r_{2},\mathbf{p}-{\bf p}^{\prime})\overline{u}(s_{2},\mathbf{p}-{\bf p}^{\prime})\right]\nonumber \\
 &  & \times f_{r_{1}s_{1}}^{(\overline{q})}(x,{\bf p}^{\prime})f_{r_{2}s_{2}}^{(q)}(x,\mathbf{p}-{\bf p}^{\prime}),\label{eq:dm2}
\end{eqnarray}
where we have changed the notation to $f_{rs}^{(q/\overline{q})}$
from $f_{rs}^{(\pm)}$. The spin density matrix element (\ref{eq:dm2})
can be put into a compact form with an explicit dependence on the
polarization vector of the quark and antiquark
\begin{eqnarray}
\rho_{\lambda_{1}\lambda_{2}}^{V}(x,{\bf p}) & = & \frac{\Delta t}{32}\int\frac{d^{3}\mathbf{p}^{\prime}}{(2\pi\hbar)^{3}}\frac{1}{E_{p^{\prime}}^{\overline{q}}E_{{\bf p}-{\bf p}^{\prime}}^{q}E_{p}^{V}}f_{\overline{q}}(x,\mathbf{p}^{\prime})f_{q}(x,\mathbf{p}-\mathbf{p}^{\prime})\nonumber \\
 &  & \times2\pi\hbar\delta\left(E_{p}^{V}-E_{p^{\prime}}^{\overline{q}}-E_{{\bf p}-{\bf p}^{\prime}}^{q}\right)\epsilon_{\alpha}^{\ast}(\lambda_{1},{\bf p})\epsilon_{\beta}(\lambda_{2},{\bf p})\nonumber \\
 &  & \times\text{Tr}\left\{ \Gamma^{\beta}\left(p^{\prime}\cdot\gamma-m_{\overline{q}}\right)\left[1+\gamma_{5}\gamma\cdot P^{\overline{q}}(x,\mathbf{p}^{\prime})\right]\Gamma^{\alpha}\right.\nonumber \\
 &  & \times\left.\left[(p-p^{\prime})\cdot\gamma+m_{q}\right]\left[1+\gamma_{5}\gamma\cdot P^{q}(x,\mathbf{p}-\mathbf{p}^{\prime})\right]\right\} ,\label{eq:spin-density-sz1-sz2}
\end{eqnarray}
where $p^{\mu}=(E_{p}^{q},\mathbf{p})$ and $p^{\prime\mu}=(E_{p^{\prime}}^{\overline{q}},\mathbf{p}^{\prime})$.
The derivation of the expression inside the trace is given in Appendix
\ref{sec:collision-kernel}. The contraction of $\epsilon_{\alpha}^{\ast}(\lambda_{1},{\bf p})$
and $\epsilon_{\beta}(\lambda_{2},{\bf p})$ with the trace can be
worked out and the result is given by Eq. (\ref{eq:i-sz1-sz2}). The
normalized $\rho_{\lambda_{1}\lambda_{2}}^{V}(x,{\bf p})$ is defined
as 
\begin{equation}
\overline{\rho}_{\lambda_{1}\lambda_{2}}^{V}(x,{\bf p})=\frac{\rho_{\lambda_{1}\lambda_{2}}^{V}(x,{\bf p})}{\mathrm{Tr}(\rho_{V})},\label{eq:norm-rho00}
\end{equation}
where $\mathrm{Tr}(\rho_{V})$ is the trace of the spin density matrix
and is evaluated using Eq. (\ref{eq:trace}) and $\rho_{\lambda_{1}\lambda_{2}}^{V}(x,{\bf p})$
is evaluated using Eq. (\ref{eq:i-sz1-sz2}).

For quarkonium vector mesons such as $\phi$ mesons with $m_{q}=m_{\overline{q}}$,
$\rho_{\lambda_{1}\lambda_{2}}^{V}(x,{\bf p})$ and $\mathrm{Tr}(\rho_{V})$
can be simplified as 
\begin{eqnarray}
\rho_{\lambda_{1}\lambda_{2}}^{V}(x,{\bf p}) & = & -\frac{\Delta t}{8}g_{V}^{2}\int\frac{d^{3}\mathbf{p}^{\prime}}{(2\pi\hbar)^{3}}\frac{1}{E_{p^{\prime}}^{\overline{q}}E_{{\bf p}-{\bf p}^{\prime}}^{q}E_{p}^{V}}f_{\overline{q}}(x,\mathbf{p}^{\prime})f_{q}(x,\mathbf{p}-\mathbf{p}^{\prime})\nonumber \\
 &  & \times2\pi\hbar\delta\left(E_{p}^{V}-E_{p^{\prime}}^{\overline{q}}-E_{{\bf p}-{\bf p}^{\prime}}^{q}\right)B^{2}(\mathbf{p}-\mathbf{p}^{\prime},\mathbf{p}^{\prime})\epsilon_{\alpha}^{\ast}(\lambda_{1},{\bf p})\epsilon_{\beta}(\lambda_{2},{\bf p})\nonumber \\
 &  & \times\left\{ \left(p^{\prime\alpha}P_{\overline{q}}^{\beta}+p^{\prime\beta}P_{\overline{q}}^{\alpha}\right)(p^{\prime}\cdot P_{q})-\left(p^{\prime\alpha}P_{q}^{\beta}+p^{\prime\beta}P_{q}^{\alpha}\right)(p\cdot P_{\overline{q}})\right.\nonumber \\
 &  & +2p^{\prime\alpha}p^{\prime\beta}(1-P_{\overline{q}}\cdot P_{q})+g^{\alpha\beta}\left[p^{\prime}\cdot p+(p^{\prime}\cdot P_{q})(p\cdot P_{\overline{q}})\right]\nonumber \\
 &  & +(p\cdot p^{\prime})\left(P_{\overline{q}}^{\alpha}P_{q}^{\beta}+P_{q}^{\alpha}P_{\overline{q}}^{\beta}-g^{\alpha\beta}P_{\overline{q}}\cdot P_{q}\right)\nonumber \\
 &  & \left.-im_{q}\varepsilon^{\alpha\beta\mu\nu}p_{\mu}(P_{\nu}^{q}+P_{\nu}^{\overline{q}})\right\} ,\label{eq:spin-density-phi}\\
\mathrm{Tr}(\rho_{V}) & = & \frac{\Delta t}{8}g_{V}^{2}\int\frac{d^{3}\mathbf{p}^{\prime}}{(2\pi\hbar)^{3}}\frac{1}{E_{p^{\prime}}^{\overline{q}}E_{{\bf p}-{\bf p}^{\prime}}^{q}E_{p}^{V}}f_{\overline{q}}(x,\mathbf{p}^{\prime})f_{q}(x,\mathbf{p}-\mathbf{p}^{\prime})\nonumber \\
 &  & \times2\pi\hbar\delta\left(E_{p}^{V}-E_{p^{\prime}}^{\overline{q}}-E_{{\bf p}-{\bf p}^{\prime}}^{q}\right)B^{2}(\mathbf{p}-\mathbf{p}^{\prime},\mathbf{p}^{\prime})\nonumber \\
 &  & \times\left[-2m_{q}^{2}(P_{\overline{q}}\cdot P_{q})+m_{V}^{2}+2m_{q}^{2}\right],\label{eq:trace-rho}
\end{eqnarray}
where we have used the short-hand notation $P_{q}\equiv P_{q}(x,\mathbf{p}-\mathbf{p}^{\prime})$
and $P_{\overline{q}}\equiv P_{\overline{q}}(x,\mathbf{p}^{\prime})$.
Equations (\ref{eq:spin-density-phi}) and (\ref{eq:trace-rho}) will
be used in the next section for evaluating spin density matrix elements
for $\phi$ mesons.

\section{Spin density matrix elements for $\phi$ mesons}

\label{sec:spin-density-phi}Now we consider the vector meson made
of a quark and its antiquark, the so-called quarkonium. For the quarkonium
vector meson such as the $\phi$ meson, the polarization distributions
in phase space in Eq. (\ref{eq:spin-density-sz1-sz2}) are given by
\citep{Becattini:2013fla,Becattini:2016gvu,Fang:2016vpj,Yang:2017sdk,Weickgenannt:2019dks}
\begin{eqnarray}
P_{s}^{\mu}(x,{\bf p}) & = & \frac{1}{4m_{s}}\epsilon^{\mu\nu\rho\sigma}\left(\omega_{\rho\sigma}+\frac{g_{\phi}}{E_{p}^{s}T_{\mathrm{eff}}}F_{\rho\sigma}^{\phi}\right)p_{\nu}\left[1-f_{s}(x,{\bf p})\right],\nonumber \\
P_{\overline{s}}^{\mu}(x,{\bf p}) & = & \frac{1}{4m_{s}}\epsilon^{\mu\nu\rho\sigma}\left(\omega_{\rho\sigma}-\frac{g_{\phi}}{E_{p}^{\overline{s}}T_{\mathrm{eff}}}F_{\rho\sigma}^{\phi}\right)p_{\nu}\left[1-f_{\overline{s}}(x,{\bf p})\right],\label{eq:polarization-ssbar}
\end{eqnarray}
where $p^{\mu}=(E_{p}^{s},{\bf p})$ and $p^{\mu}=(E_{p}^{\overline{s}},{\bf p})$
denote the four-momenta of the strange quark $s$ and antiquark $\overline{s}$
respectively, with $E_{p}^{s}=E_{p}^{\overline{s}}=\sqrt{|\mathbf{p}|^{2}+m_{s}^{2}}$
and $m_{\overline{s}}=m_{s}$. We have assumed that $s$ and $\overline{s}$
are polarized by the thermal vorticity (tensor) field $\omega_{\rho\sigma}=(1/2)[\partial_{\rho}(\beta u_{\sigma})-\partial_{\sigma}(\beta u_{\rho})]$
and $\phi$ field strength tensor $F_{\rho\sigma}^{\phi}=\partial_{\rho}A_{\sigma}^{\phi}-\partial_{\sigma}A_{\rho}^{\phi}$
\citep{Sheng:2019kmk}, where $u_{\sigma}$ is the fluid velocity,
$\beta=1/T_{\mathrm{eff}}$ is the inverse effective temperature,
and $A_{\sigma}^{\phi}$ is the vector potential of the $\phi$ field.
Note that in some literature the definition of $\omega_{\rho\sigma}$
may differ by a sign \citep{Becattini:2013fla,Becattini:2016gvu,Huang:2020dtn}.
In Eq. (\ref{eq:polarization-ssbar}) $f_{s}(x,{\bf p})$ and $f_{\overline{s}}(x,{\bf p})$
are unpolarized phase space distributions of $s$ and $\overline{s}$
respectively and given by the Fermi-Dirac distribution 
\begin{equation}
f_{s/\overline{s}}(x,{\bf p})=\frac{1}{1+\exp(\beta E_{p}^{s/\overline{s}}\mp\mu_{s})}
\end{equation}
where $\mu_{s}$ is the chemical potential for $s$ ($-\mu_{s}$ for
$\overline{s}$). In most cases $f_{s/\overline{s}}$ are negligible
relative to 1 in $P_{s/\overline{s}}^{\mu}$ in Eq. (\ref{eq:polarization-ssbar}).
The spin-field coupling in (\ref{eq:polarization-ssbar}) can be derived
from the Wigner functions for massive fermions \citep{Weickgenannt:2019dks}
and has a clear physical meaning: one contribution is from the magnetic
field through the magnetic moment and the other contribution from
the electric field through the spin-orbit coupling, the former is
always there while the latter is only present for moving fermions.
The mean field effects of vector mesons have been studied in the context
of spin polarization of $\Lambda$ hyperons \citep{Csernai:2018yok}
and different elliptic flows between hadrons of some species and their
antiparticles \citep{Xu:2012gf} in heavy-ion collisions.

The spin direction four-vector for the $\phi$ meson is given by 
\begin{equation}
\epsilon^{\mu}(\lambda,{\bf p})=\left(\frac{{\bf p}\cdot\boldsymbol{\epsilon}_{\lambda}}{m_{\phi}},\boldsymbol{\epsilon}_{\lambda}+\frac{{\bf p}\cdot\boldsymbol{\epsilon}_{\lambda}}{m_{\phi}(E_{p}^{\phi}+m_{\phi})}{\bf p}\right),
\end{equation}
where $E_{p}^{\phi}\equiv\sqrt{m_{\phi}^{2}+{\bf p}^{2}}$ is the
energy of the $\phi$ meson, $\lambda=0,\pm1$ denotes the spin states,
and $\boldsymbol{\epsilon}_{\lambda}$ denotes the three-vector of
the spin state (spin vector) in the $\phi$ meson's rest frame. In
order to calculate the spin alignment along the direction of the global
orbital angular momentum (the $y$-direction) in heavy-ion collisions,
we choose the $y$-direction as the spin quantization direction. So
the corresponding spin vectors are 
\begin{eqnarray}
\boldsymbol{\epsilon}_{0} & = & \left(0,1,0\right),\nonumber \\
\boldsymbol{\epsilon}_{+1} & = & -\frac{1}{\sqrt{2}}\left(i,0,1\right),\nonumber \\
\boldsymbol{\epsilon}_{-1} & = & \frac{1}{\sqrt{2}}\left(-i,0,1\right).
\end{eqnarray}
The 00-component of the spin density matrix is what can be measured
in experiments which concerns the real vector $\boldsymbol{\epsilon}_{0}$
satisfying $\boldsymbol{\epsilon}_{0}=\boldsymbol{\epsilon}_{0}^{*}$.

Substituting Eq. (\ref{eq:polarization-ssbar}) into Eq. (\ref{eq:spin-density-sz1-sz2}),
we obtain 
\begin{eqnarray}
\rho_{\lambda_{1}\lambda_{2}}^{\phi} & = & \rho_{\lambda_{1}\lambda_{2}}^{\phi}(0)+\rho_{\lambda_{1}\lambda_{2}}^{\phi}(\omega^{1})+\rho_{\lambda_{1}\lambda_{2}}^{\phi}(F_{\phi}^{1})\nonumber \\
 &  & +\rho_{\lambda_{1}\lambda_{2}}^{\phi}(\omega^{2})+\rho_{\lambda_{1}\lambda_{2}}^{\phi}(F_{\phi}^{2}),\label{eq:rho-00-omega-f}
\end{eqnarray}
where $\omega^{i}$ and $F_{\phi}^{i}$ with $i=0,1,2$ denote the
zeroth, first, and second order terms in the vorticity and $\phi$
field respectively. The zeroth order term $\rho_{\lambda_{1}\lambda_{2}}^{\phi}(0)$
represents the unpolarized contribution. In (\ref{eq:rho-00-omega-f}),
we neglected mixing terms of $\omega_{\mu\nu}$ and $F_{\mu\nu}^{\phi}$
since we assume that there is no correlation between them in space-time
so these terms are vanishing after taking a space-time average of
$\rho_{\lambda_{1}\lambda_{2}}^{\phi}$. For $\lambda_{1}=\lambda_{2}=0$,
$\epsilon_{\alpha}^{\ast}(0,{\bf p})\epsilon_{\beta}(0,{\bf p})=\epsilon_{\alpha}(0,{\bf p})\epsilon_{\beta}(0,{\bf p})$
is symmetric in $\alpha$ and $\beta$, then one can verify that the
first order terms $\rho_{00}^{\phi}(\omega^{1})$ and $\rho_{00}^{\phi}(F_{\phi}^{1})$
are vanishing. The zeroth order term $\rho_{00}^{\phi}(0)$ is given
by 
\begin{eqnarray}
\rho_{00}^{\phi}(0) & = & \frac{\Delta t}{8}g_{\phi}^{2}\int\frac{d^{3}\mathbf{p}^{\prime}}{(2\pi\hbar)^{3}}\frac{1}{E_{p^{\prime}}^{\overline{s}}E_{{\bf p}-{\bf p}^{\prime}}^{s}E_{p}^{\phi}}f_{\overline{s}}(\mathbf{p}^{\prime})f_{s}(\mathbf{p}-\mathbf{p}^{\prime})B^{2}(\mathbf{p}-\mathbf{p}^{\prime},\mathbf{p}^{\prime})\nonumber \\
 &  & \times2\pi\hbar\delta\left(E_{p}^{\phi}-E_{p^{\prime}}^{\overline{s}}-E_{{\bf p}-{\bf p}^{\prime}}^{s}\right)\left\{ (p^{\prime}\cdot p)-2[p^{\prime}\cdot\epsilon(0,{\bf p})]^{2}\right\} ,\label{eq:rho00-0}
\end{eqnarray}
where we have used the second relation of Eq. (\ref{eq:polar-vector-rel}).
The second order terms $\rho_{\lambda_{1}\lambda_{2}}^{\phi}(\omega^{2})$
and $\rho_{\lambda_{1}\lambda_{2}}^{\phi}(F_{\phi}^{2})$ read 
\begin{eqnarray}
\rho_{\lambda_{1}\lambda_{2}}^{\phi}(\omega^{2}) & \approx & -\frac{\Delta t}{32}\frac{1}{4m_{s}^{2}}g_{\phi}^{2}\int\frac{d^{3}\mathbf{p}^{\prime}}{(2\pi\hbar)^{3}}\frac{1}{E_{p^{\prime}}^{\overline{s}}E_{{\bf p}-{\bf p}^{\prime}}^{s}E_{p}^{\phi}}B^{2}(\mathbf{p}-\mathbf{p}^{\prime},\mathbf{p}^{\prime})\nonumber \\
 &  & \times f_{\overline{s}}(\mathbf{p}^{\prime})f_{s}(\mathbf{p}-\mathbf{p}^{\prime})2\pi\hbar\delta\left(E_{p}^{\phi}-E_{p^{\prime}}^{\overline{s}}-E_{{\bf p}-{\bf p}^{\prime}}^{s}\right)\nonumber \\
 &  & \times\epsilon_{\alpha}^{\ast}(\lambda_{1},{\bf p})\epsilon_{\beta}(\lambda_{2},{\bf p})\widetilde{\omega}_{\rho\xi}(x)\widetilde{\omega}_{\sigma\gamma}(x)p^{\prime\xi}(p-p^{\prime})^{\gamma}\nonumber \\
 &  & \times\text{Tr}\left\{ \gamma^{\beta}\left(p^{\prime}\cdot\gamma+m_{s}\right)\gamma^{\rho}\gamma^{\alpha}\left[(p-p^{\prime})\cdot\gamma+m_{s}\right]\gamma^{\sigma}\right\} ,\label{eq:rho-00-omega}
\end{eqnarray}
and 
\begin{eqnarray}
\rho_{\lambda_{1}\lambda_{2}}^{\phi}(F_{\phi}^{2}) & \approx & \frac{\Delta t}{32}\frac{1}{4m_{s}^{2}T_{\mathrm{eff}}^{2}}g_{\phi}^{4}\int\frac{d^{3}\mathbf{p}^{\prime}}{(2\pi\hbar)^{3}}\frac{1}{(E_{p^{\prime}}^{\overline{s}})^{2}(E_{{\bf p}-{\bf p}^{\prime}}^{s})^{2}E_{p}^{\phi}}B^{2}(\mathbf{p}-\mathbf{p}^{\prime},\mathbf{p}^{\prime})\nonumber \\
 &  & \times f_{\overline{s}}(\mathbf{p}^{\prime})f_{s}(\mathbf{p}-\mathbf{p}^{\prime})2\pi\hbar\delta\left(E_{p}^{\phi}-E_{p^{\prime}}^{\overline{s}}-E_{{\bf p}-{\bf p}^{\prime}}^{s}\right)\nonumber \\
 &  & \times\epsilon_{\alpha}^{\ast}(\lambda_{1},{\bf p})\epsilon_{\beta}(\lambda_{2},{\bf p})\widetilde{F}_{\rho\xi}^{\phi}(x)\widetilde{F}_{\sigma\gamma}^{\phi}(x)p^{\prime\xi}(p-p^{\prime})^{\gamma}\nonumber \\
 &  & \times\text{Tr}\left\{ \gamma^{\beta}\left(p^{\prime}\cdot\gamma+m_{s}\right)\gamma^{\rho}\gamma^{\alpha}\left[(p-p^{\prime})\cdot\gamma+m_{s}\right]\gamma^{\sigma}\right\} .\label{eq:rho-f2}
\end{eqnarray}
In Eqs. (\ref{eq:rho-00-omega}) and (\ref{eq:rho-f2}) we have used
$\widetilde{\omega}_{\rho\xi}=(1/2)\epsilon_{\rho\xi\alpha\beta}\omega^{\alpha\beta}$,
$\widetilde{F}_{\rho\xi}^{\phi}=(1/2)\epsilon_{\rho\xi\alpha\beta}F_{\phi}^{\alpha\beta}$,
and neglected $f_{s/\overline{s}}$ relative to 1 in $P_{s/\overline{s}}^{\mu}$.
The tensor part of $\rho_{\lambda_{1}\lambda_{2}}^{\phi}(\omega^{2})$
and $\rho_{\lambda_{1}\lambda_{2}}^{\phi}(F_{\phi}^{2})$ that is
contracted with $\epsilon_{\alpha}^{\ast}\epsilon_{\beta}\widetilde{\omega}_{\rho\xi}\widetilde{\omega}_{\sigma\gamma}$
and $\epsilon_{\alpha}^{\ast}\epsilon_{\beta}\widetilde{F}_{\rho\xi}^{\phi}\widetilde{F}_{\sigma\gamma}^{\phi}$
respectively can be evaluated as 
\begin{eqnarray}
I^{\alpha\beta;\rho\xi;\sigma\gamma} & = & p^{\prime\xi}(p-p^{\prime})^{\gamma}\text{Tr}\left\{ \gamma^{\beta}\left(p^{\prime}\cdot\gamma+m_{s}\right)\gamma^{\rho}\gamma^{\alpha}\left[(p-p^{\prime})\cdot\gamma+m_{s}\right]\gamma^{\sigma}\right\} \nonumber \\
 & = & 2p^{\prime\xi}p^{\gamma}\left[m_{\phi}^{2}\left(g^{\beta\rho}g^{\alpha\sigma}-g^{\alpha\beta}g^{\rho\sigma}+g^{\beta\sigma}g^{\alpha\rho}\right)\right.\nonumber \\
 &  & +2p^{\rho}\left(g^{\alpha\beta}p^{\prime\sigma}-g^{\alpha\sigma}p^{\prime\beta}-g^{\beta\sigma}p^{\prime\alpha}\right)\nonumber \\
 &  & \left.+2\left(g^{\beta\rho}p^{\prime\alpha}p^{\prime\sigma}+g^{\alpha\rho}p^{\prime\beta}p^{\prime\sigma}-2g^{\rho\sigma}p^{\prime\alpha}p^{\prime\beta}\right)\right]\nonumber \\
 &  & -2p^{\prime\xi}p^{\prime\gamma}\left[m_{\phi}^{2}\left(g^{\beta\rho}g^{\alpha\sigma}-g^{\alpha\beta}g^{\rho\sigma}+g^{\beta\sigma}g^{\alpha\rho}\right)\right.\nonumber \\
 &  & \left.-2p^{\rho}\left(g^{\alpha\sigma}p^{\prime\beta}+g^{\beta\sigma}p^{\prime\alpha}\right)-4g^{\rho\sigma}p^{\prime\alpha}p^{\prime\beta}\right].\label{eq:eval-trace}
\end{eqnarray}
With the above tensor and the quantity inside the curly brackets in
(\ref{eq:rho00-0}), $\rho_{00}^{\phi}(0)$, $\rho_{\lambda_{1}\lambda_{2}}^{\phi}(\omega^{2})$
and $\rho_{\lambda_{1}\lambda_{2}}^{\phi}(F^{2})$ involve following
moments of momenta 
\begin{eqnarray}
\left\{ I_{0},I_{0}^{\mu},I_{0}^{\mu\nu},I_{0}^{\mu\nu\rho},I_{0}^{\mu\nu\rho\sigma}\right\}  & = & \int\frac{d^{3}\mathbf{p}^{\prime}}{(2\pi\hbar)^{3}}\frac{1}{E_{p^{\prime}}^{\overline{s}}E_{{\bf p}-{\bf p}^{\prime}}^{s}}B^{2}(\mathbf{p}-\mathbf{p}^{\prime},\mathbf{p}^{\prime})\nonumber \\
 &  & \times f_{\overline{s}}(\mathbf{p}^{\prime})f_{s}(\mathbf{p}-\mathbf{p}^{\prime})2\pi\hbar\delta\left(E_{p}^{\phi}-E_{p^{\prime}}^{\overline{s}}-E_{{\bf p}-{\bf p}^{\prime}}^{s}\right)\nonumber \\
 &  & \times\left\{ 1,p^{\prime\mu},p^{\prime\mu}p^{\prime\nu},p^{\prime\mu}p^{\prime\nu}p^{\prime\rho},p^{\prime\mu}p^{\prime\nu}p^{\prime\rho}p^{\prime\sigma}\right\} ,\label{eq:i-omega}\\
\left\{ I_{F}^{\mu},I_{F}^{\mu\nu},I_{F}^{\mu\nu\rho},I_{F}^{\mu\nu\rho\sigma}\right\}  & = & \int\frac{d^{3}\mathbf{p}^{\prime}}{(2\pi\hbar)^{3}}\frac{1}{(E_{p^{\prime}}^{\overline{s}})^{2}(E_{{\bf p}-{\bf p}^{\prime}}^{s})^{2}}B^{2}(\mathbf{p}-\mathbf{p}^{\prime},\mathbf{p}^{\prime})\nonumber \\
 &  & \times f_{\overline{s}}(\mathbf{p}^{\prime})f_{s}(\mathbf{p}-\mathbf{p}^{\prime})2\pi\hbar\delta\left(E_{p}^{\phi}-E_{p^{\prime}}^{\overline{s}}-E_{{\bf p}-{\bf p}^{\prime}}^{s}\right)\nonumber \\
 &  & \times\left\{ p^{\prime\mu},p^{\prime\mu}p^{\prime\nu},p^{\prime\mu}p^{\prime\nu}p^{\prime\rho},p^{\prime\mu}p^{\prime\nu}p^{\prime\rho}p^{\prime\sigma}\right\} .\label{eq:i-f-tensor}
\end{eqnarray}
The tensors in (\ref{eq:i-omega}) with the subscript '0' are those
in $\rho_{00}^{\phi}(0)$ and $\rho_{\lambda_{1}\lambda_{2}}^{\phi}(\omega^{2})$,
and the tensors in (\ref{eq:i-f-tensor}) with the subscript 'F' are
those in $\rho_{\lambda_{1}\lambda_{2}}^{\phi}(F_{\phi}^{2})$. The
difference between Eq. (\ref{eq:i-omega}) and Eq. (\ref{eq:i-f-tensor})
is in the powers of $E_{p^{\prime}}^{\overline{s}}$ and $E_{{\bf p}-{\bf p}^{\prime}}^{s}$
in the denominators. Note that all above tensors with 2 or more indices
are symmetric with respect to the interchange of any two indices.

Using Eqs. (\ref{eq:eval-trace})-(\ref{eq:i-f-tensor}), the zeroth
and second order terms of the spin density matrix in (\ref{eq:rho00-0}),
(\ref{eq:rho-00-omega}) and (\ref{eq:rho-f2}) can be expressed in
terms of moments of momenta
\begin{equation}
\rho_{00}^{\phi}(0)=\frac{\Delta t}{16}g_{\phi}^{2}m_{\phi}^{2}\frac{1}{E_{p}^{\phi}}I_{0}\left[1-4\epsilon_{\alpha}(0,{\bf p})\epsilon_{\beta}(0,{\bf p})\frac{I_{0}^{\alpha\beta}}{m_{\phi}^{2}I_{0}}\right],\label{eq:rho-00-0th}
\end{equation}
\begin{eqnarray}
\rho_{00}^{\phi}(\omega^{2}) & = & -\frac{\Delta t}{64}\frac{g_{\phi}^{2}}{m_{s}^{2}}\frac{1}{E_{p}^{\phi}}\epsilon_{\alpha}(0,{\bf p})\epsilon_{\beta}(0,{\bf p})\widetilde{\omega}_{\rho\xi}(x)\widetilde{\omega}_{\sigma\gamma}(x)\nonumber \\
 &  & \times\left[p^{\gamma}m_{\phi}^{2}\left(g^{\beta\rho}g^{\alpha\sigma}-g^{\alpha\beta}g^{\rho\sigma}+g^{\beta\sigma}g^{\alpha\rho}\right)I_{0}^{\xi}\right.\nonumber \\
 &  & +2p^{\gamma}p^{\rho}\left(g^{\alpha\beta}I_{0}^{\xi\sigma}-g^{\alpha\sigma}I_{0}^{\xi\beta}-g^{\beta\sigma}I_{0}^{\xi\alpha}\right)\nonumber \\
 &  & +2p^{\gamma}\left(g^{\beta\rho}I_{0}^{\xi\sigma\alpha}+I_{0}^{\xi\sigma\beta}g^{\alpha\rho}-2g^{\rho\sigma}I_{0}^{\xi\alpha\beta}\right)\nonumber \\
 &  & -m_{\phi}^{2}\left(g^{\beta\rho}g^{\alpha\sigma}-g^{\alpha\beta}g^{\rho\sigma}+g^{\beta\sigma}g^{\alpha\rho}\right)I_{0}^{\xi\gamma}\nonumber \\
 &  & \left.+2p^{\rho}\left(g^{\alpha\sigma}I_{0}^{\xi\beta\gamma}+g^{\beta\sigma}I_{0}^{\xi\alpha\gamma}\right)+4g^{\rho\sigma}I_{0}^{\xi\alpha\beta\gamma}\right],\label{eq:rho-00-omega-2nd}
\end{eqnarray}
\begin{eqnarray}
\rho_{00}^{\phi}(F_{\phi}^{2}) & = & \frac{\Delta t}{64}\frac{g_{\phi}^{4}}{m_{s}^{2}T_{\mathrm{eff}}^{2}}\frac{1}{E_{p}^{\phi}}\epsilon_{\alpha}(0,{\bf p})\epsilon_{\beta}(0,{\bf p})\widetilde{F}_{\rho\xi}^{\phi}(x)\widetilde{F}_{\sigma\gamma}^{\phi}(x)\nonumber \\
 &  & \times\left[p^{\gamma}m_{\phi}^{2}\left(g^{\beta\rho}g^{\alpha\sigma}-g^{\alpha\beta}g^{\rho\sigma}+g^{\beta\sigma}g^{\alpha\rho}\right)I_{F}^{\xi}\right.\nonumber \\
 &  & +2p^{\gamma}p^{\rho}\left(g^{\alpha\beta}I_{F}^{\xi\sigma}-g^{\alpha\sigma}I_{F}^{\xi\beta}-g^{\beta\sigma}I_{F}^{\xi\alpha}\right)\nonumber \\
 &  & +2p^{\gamma}\left(g^{\beta\rho}I_{F}^{\xi\sigma\alpha}+I_{F}^{\xi\sigma\beta}g^{\alpha\rho}-2g^{\rho\sigma}I_{F}^{\xi\alpha\beta}\right)\nonumber \\
 &  & -m_{\phi}^{2}\left(g^{\beta\rho}g^{\alpha\sigma}-g^{\alpha\beta}g^{\rho\sigma}+g^{\beta\sigma}g^{\alpha\rho}\right)I_{F}^{\xi\gamma}\nonumber \\
 &  & \left.+2p^{\rho}\left(g^{\alpha\sigma}I_{F}^{\xi\beta\gamma}+g^{\beta\sigma}I_{F}^{\xi\alpha\gamma}\right)+4g^{\rho\sigma}I_{F}^{\xi\alpha\beta\gamma}\right].\label{eq:rho00-f-2nd}
\end{eqnarray}
From Eq. (\ref{eq:trace-rho}), the trace of the spin density matrix
for the $\phi$ meson reads
\begin{eqnarray}
\mathrm{Tr}(\rho_{\phi}) & = & \frac{\Delta t}{8}g_{\phi}^{2}(m_{\phi}^{2}+2m_{s}^{2})\frac{1}{E_{p}^{\phi}}I_{0}\nonumber \\
 &  & \times\left[1-\frac{1}{2(m_{\phi}^{2}+2m_{s}^{2})}\widetilde{\omega}_{\rho\xi}(x)\widetilde{\omega}_{\sigma\gamma}(x)g^{\rho\sigma}\frac{1}{I_{0}}\left(p^{\gamma}I_{0}^{\xi}-I_{0}^{\xi\gamma}\right)\right.\nonumber \\
 &  & \left.+\frac{g_{\phi}^{2}}{2(m_{\phi}^{2}+2m_{s}^{2})T_{\mathrm{eff}}^{2}}\widetilde{F}_{\rho\xi}^{\phi}(x)\widetilde{F}_{\sigma\gamma}^{\phi}(x)g^{\rho\sigma}\frac{1}{I_{0}}\left(p^{\gamma}I_{F}^{\xi}-I_{F}^{\xi\gamma}\right)\right].\label{eq:trace-rho00}
\end{eqnarray}
Here we have neglected mixing terms of $\omega_{\mu\nu}$ and $F_{\mu\nu}^{\phi}$
since we assume that there is no correlation in space-time between
them.

From Eqs. (\ref{eq:rho-00-0th})-(\ref{eq:trace-rho00}) we obtain
the 00-component of the normalized spin density matrix for the $\phi$
meson defined in (\ref{eq:norm-rho00}) 
\begin{equation}
\overline{\rho}_{00}^{\phi}(x,{\bf p})=c_{0}({\bf p})+c_{\omega}(x,{\bf p})+c_{F}(x,{\bf p}),\label{eq:av-rho00}
\end{equation}
where $c_{0}$, $c_{\omega}$ and $c_{F}$ are given by 
\begin{equation}
c_{0}({\bf p})=\frac{m_{\phi}^{2}}{2(m_{\phi}^{2}+2m_{s}^{2})}\left[1-4\epsilon_{\alpha}(0,{\bf p})\epsilon_{\beta}(0,{\bf p})\frac{I_{0}^{\alpha\beta}}{m_{\phi}^{2}I_{0}}\right],\label{eq:c0-p}
\end{equation}
\begin{eqnarray}
c_{\omega}(x,{\bf p}) & = & -\frac{1}{8m_{s}^{2}(m_{\phi}^{2}+2m_{s}^{2})}\epsilon_{\alpha}(0,{\bf p})\epsilon_{\beta}(0,{\bf p})\widetilde{\omega}_{\rho\xi}(x)\widetilde{\omega}_{\sigma\gamma}(x)\nonumber \\
 &  & \times\frac{1}{I_{0}}\left[p^{\gamma}m_{\phi}^{2}\left(g^{\beta\rho}g^{\alpha\sigma}-g^{\alpha\beta}g^{\rho\sigma}+g^{\beta\sigma}g^{\alpha\rho}\right)I_{0}^{\xi}\right.\nonumber \\
 &  & +2p^{\gamma}p^{\rho}\left(g^{\alpha\beta}I_{0}^{\xi\sigma}-g^{\alpha\sigma}I_{0}^{\xi\beta}-g^{\beta\sigma}I_{0}^{\xi\alpha}\right)\nonumber \\
 &  & +2p^{\gamma}\left(g^{\beta\rho}I_{0}^{\xi\sigma\alpha}+I_{0}^{\xi\sigma\beta}g^{\alpha\rho}-2g^{\rho\sigma}I_{0}^{\xi\alpha\beta}\right)\nonumber \\
 &  & -m_{\phi}^{2}\left(g^{\beta\rho}g^{\alpha\sigma}-g^{\alpha\beta}g^{\rho\sigma}+g^{\beta\sigma}g^{\alpha\rho}\right)I_{0}^{\xi\gamma}\nonumber \\
 &  & \left.+2p^{\rho}\left(g^{\alpha\sigma}I_{0}^{\xi\beta\gamma}+g^{\beta\sigma}I_{0}^{\xi\alpha\gamma}\right)+4g^{\rho\sigma}I_{0}^{\xi\alpha\beta\gamma}\right]\nonumber \\
 &  & +\frac{c_{0}({\bf p})}{2(m_{\phi}^{2}+2m_{s}^{2})}\widetilde{\omega}_{\rho\xi}(x)\widetilde{\omega}_{\sigma\gamma}(x)g^{\rho\sigma}\frac{1}{I_{0}}\left(p^{\gamma}I_{0}^{\xi}-I_{0}^{\xi\gamma}\right),\label{eq:c-phi-av}
\end{eqnarray}
and 
\begin{eqnarray}
c_{F}(x,{\bf p}) & = & \frac{1}{8m_{s}^{2}(m_{\phi}^{2}+2m_{s}^{2})}\frac{g_{\phi}^{2}}{T_{\mathrm{eff}}^{2}}\epsilon_{\alpha}(0,{\bf p})\epsilon_{\beta}(0,{\bf p})\widetilde{F}_{\rho\xi}^{\phi}(x)\widetilde{F}_{\sigma\gamma}^{\phi}(x)\nonumber \\
 &  & \times\frac{1}{I_{0}}\left[p^{\gamma}m_{\phi}^{2}\left(g^{\beta\rho}g^{\alpha\sigma}-g^{\alpha\beta}g^{\rho\sigma}+g^{\beta\sigma}g^{\alpha\rho}\right)I_{F}^{\xi}\right.\nonumber \\
 &  & +2p^{\gamma}p^{\rho}\left(g^{\alpha\beta}I_{F}^{\xi\sigma}-g^{\alpha\sigma}I_{F}^{\xi\beta}-g^{\beta\sigma}I_{F}^{\xi\alpha}\right)\nonumber \\
 &  & +2p^{\gamma}\left(g^{\beta\rho}I_{F}^{\xi\sigma\alpha}+I_{F}^{\xi\sigma\beta}g^{\alpha\rho}-2g^{\rho\sigma}I_{F}^{\xi\alpha\beta}\right)\nonumber \\
 &  & -m_{\phi}^{2}\left(g^{\beta\rho}g^{\alpha\sigma}-g^{\alpha\beta}g^{\rho\sigma}+g^{\beta\sigma}g^{\alpha\rho}\right)I_{F}^{\xi\gamma}\nonumber \\
 &  & \left.+2p^{\rho}\left(g^{\alpha\sigma}I_{F}^{\xi\beta\gamma}+g^{\beta\sigma}I_{F}^{\xi\alpha\gamma}\right)+4g^{\rho\sigma}I_{F}^{\xi\alpha\beta\gamma}\right]\nonumber \\
 &  & -\frac{g_{\phi}^{2}c_{0}({\bf p})}{2(m_{\phi}^{2}+2m_{s}^{2})T_{\mathrm{eff}}^{2}}\widetilde{F}_{\rho\xi}^{\phi}(x)\widetilde{F}_{\sigma\gamma}^{\phi}(x)g^{\rho\sigma}\frac{1}{I_{0}}\left(p^{\gamma}I_{F}^{\xi}-I_{F}^{\xi\gamma}\right).\label{eq:cf-av}
\end{eqnarray}
We see in Eqs. (\ref{eq:c0-p})-(\ref{eq:cf-av}) that the momentum
moments always come with the factor $1/I_{0}$, so they can be understood
as normalized moments by $I_{0}$, a kind of momentum averages.

\begin{table}
\begin{tabular}{|c|c|c|c|c|c|c|c|c|c|}
\hline 
 & $I^{\mu}$ & $I^{00}$ & $I^{aa}$ & $I^{000}$ & $I^{0aa}$ & $I^{0000}$ & $I^{00aa}$ & $I^{aabb}$ & $I^{aaaa}$\tabularnewline
\hline 
$\omega$ & $(m_{\phi}/2)g^{\mu0}$ & $m_{\phi}^{2}/4$ & $d_{0}m_{\phi}^{2}/4$ & $m_{\phi}^{3}/8$ & $d_{0}m_{\phi}^{3}/8$ & $m_{\phi}^{4}/16$ & $d_{0}m_{\phi}^{4}/16$ & $d_{0}^{2}m_{\phi}^{4}/80$ & $3d_{0}^{2}m_{\phi}^{4}/80$\tabularnewline
\hline 
$F_{\phi}$ & $(2/m_{\phi})g^{\mu0}$ & $1$ & $d_{0}$ & $m_{\phi}/2$ & $d_{0}m_{\phi}/2$ & $m_{\phi}^{2}/4$ & $d_{0}m_{\phi}^{2}/4$ & $d_{0}^{2}m_{\phi}^{2}/20$ & $3d_{0}^{2}m_{\phi}^{2}/20$\tabularnewline
\hline 
\end{tabular}

\caption{\label{tab:momentum-moment}All nonvanishing moments of momenta normalized
by $I_{0}$ in $\overline{\rho}_{00}^{\phi}$ from contributions of
the vorticity and the $\phi$ field, which are evaluated in the rest
frame of the vector meson. Note that $I$ represents either $I_{0}$
or $I_{F}$. The definition for some quantities are $I^{aa}\equiv I^{11}+I^{22}+I^{33}$,
$I^{0aa}\equiv I^{011}+I^{022}+I^{033}$, $I^{00aa}\equiv I^{0011}+I^{0022}+I^{0033}$,
$I^{aabb}\equiv I^{1122}+I^{2233}+I^{3311}$, and $I^{aaaa}\equiv I^{1111}+I^{2222}+I^{3333}$.
The constant $d_{0}$ is defined as $d_{0}\equiv1-4m_{s}^{2}/m_{\phi}^{2}$.}
\end{table}

We see in Eqs. (\ref{eq:c0-p})-(\ref{eq:cf-av}) that $c_{0}$, $c_{\omega}$
and $c_{F}$ are all Lorentz scalars, so it is convenient to evaluate
them in the rest frame of the vector meson. All nonvanishing moments
of momenta in Eqs. (\ref{eq:c0-p})-(\ref{eq:cf-av}) that are evaluated
in the rest frame of the vector meson are listed in Table \ref{tab:momentum-moment}.
Finally the result for $\overline{\rho}_{00}^{\phi}$ is 
\begin{eqnarray}
\overline{\rho}_{00}^{\phi}(x,{\bf p}) & \approx & \frac{1}{3}+C_{1}\left[\frac{1}{3}\left(\boldsymbol{\omega}^{\prime}\cdot\boldsymbol{\omega}^{\prime}-\frac{4g_{\phi}^{2}}{m_{\phi}^{2}T_{\mathrm{eff}}^{2}}{\bf B}_{\phi}^{\prime}\cdot{\bf B}_{\phi}^{\prime}\right)-(\boldsymbol{\epsilon}_{0}\cdot\boldsymbol{\omega}^{\prime})^{2}+\frac{4g_{\phi}^{2}}{m_{\phi}^{2}T_{\mathrm{eff}}^{2}}\left(\boldsymbol{\epsilon}_{0}\cdot{\bf B}_{\phi}^{\prime}\right)^{2}\right]\nonumber \\
 &  & +C_{2}\left[\frac{1}{3}\left(\boldsymbol{\varepsilon}^{\prime}\cdot\boldsymbol{\varepsilon}^{\prime}-\frac{4g_{\phi}^{2}}{m_{\phi}^{2}T_{\mathrm{eff}}^{2}}{\bf E}_{\phi}^{\prime}\cdot{\bf E}_{\phi}^{\prime}\right)-(\boldsymbol{\epsilon}_{0}\cdot\boldsymbol{\varepsilon}^{\prime})^{2}+\frac{4g_{\phi}^{2}}{m_{\phi}^{2}T_{\mathrm{eff}}^{2}}(\boldsymbol{\epsilon}_{0}\cdot{\bf E}_{\phi}^{\prime})^{2}\right],\label{eq:av-rho00-square}
\end{eqnarray}
where the fields with primes are in the rest frame of the vector meson,
$\boldsymbol{\varepsilon}$ and $\boldsymbol{\omega}$ denote the
electric and magnetic part of the vorticity tensor $\omega^{\mu\nu}$
respectively, ${\bf E}_{\phi}$ and ${\bf B}_{\phi}$ denote the electric
and magnetic part of the $\phi$ field tensor $F_{\phi}^{\mu\nu}$
respectively, and $C_{1}$ and $C_{2}$ are two coefficients depending
on masses of the quark and vector meson defined as 
\begin{eqnarray}
C_{1} & = & \frac{8m_{s}^{4}+16m_{s}^{2}m_{\phi}^{2}+3m_{\phi}^{4}}{120m_{s}^{2}(m_{\phi}^{2}+2m_{s}^{2})},\nonumber \\
C_{2} & = & \frac{8m_{s}^{4}-14m_{s}^{2}m_{\phi}^{2}+3m_{\phi}^{4}}{120m_{s}^{2}(m_{\phi}^{2}+2m_{s}^{2})}.\label{eq:constant-mass}
\end{eqnarray}
The result for $\overline{\rho}_{00}^{\phi}(x,{\bf p})$ in Eq. (\ref{eq:av-rho00-square})
is rigorous and remarkable since all contributions are in squares
of the fields. This is a clear piece of evidence that there exists
in the $\phi$ meson an exact correlation between the strong force
field coupled to the $s$ quark and that coupled to the $\overline{s}$
quark. This feature makes $\rho_{00}$ for quarkonium vector mesons
very different from that for other vector mesons carrying net charges
or flavors.

One can approximate $\overline{\rho}_{00}^{\phi}$ by expanding $C_{1}$
and $C_{2}$ in terms of the average quark momentum inside the vector
meson as 
\begin{eqnarray}
C_{1} & \approx & \frac{1}{6}+\frac{1}{9}d_{0}+O(d_{0}^{2}),\nonumber \\
C_{2} & \approx & \frac{1}{18}d_{0}+O(d_{0}^{2}),
\end{eqnarray}
with $d_{0}\equiv1-4m_{s}^{2}/m_{\phi}^{2}$, the result is 
\begin{eqnarray}
\overline{\rho}_{00}^{\phi}(x,{\bf p}) & \approx & \frac{1}{3}+\left(\frac{1}{6}+\frac{1}{9}d_{0}\right)\left\{ \frac{1}{3}\left(\boldsymbol{\omega}^{\prime}\cdot\boldsymbol{\omega}^{\prime}-\frac{4g_{\phi}^{2}}{m_{\phi}^{2}T_{\mathrm{eff}}^{2}}{\bf B}_{\phi}^{\prime}\cdot{\bf B}_{\phi}^{\prime}\right)\right.\nonumber \\
 &  & \left.-(\boldsymbol{\epsilon}_{0}\cdot\boldsymbol{\omega}^{\prime})^{2}+\frac{4g_{\phi}^{2}}{m_{\phi}^{2}T_{\mathrm{eff}}^{2}}\left(\boldsymbol{\epsilon}_{0}\cdot{\bf B}_{\phi}^{\prime}\right)^{2}\right\} \nonumber \\
 &  & +\frac{1}{18}d_{0}\left\{ \frac{1}{3}\left(\boldsymbol{\varepsilon}^{\prime}\cdot\boldsymbol{\varepsilon}^{\prime}-\frac{4g_{\phi}^{2}}{m_{\phi}^{2}T_{\mathrm{eff}}^{2}}{\bf E}_{\phi}^{\prime}\cdot{\bf E}_{\phi}^{\prime}\right)\right.\nonumber \\
 &  & \left.-(\boldsymbol{\epsilon}_{0}\cdot\boldsymbol{\varepsilon}^{\prime})^{2}+\frac{4g_{\phi}^{2}}{m_{\phi}^{2}T_{\mathrm{eff}}^{2}}(\boldsymbol{\epsilon}_{0}\cdot{\bf E}_{\phi}^{\prime})^{2}\right\} +O(d_{0}^{2}).\label{eq:rho00-app-eb}
\end{eqnarray}
The above result can be compared with that in the nonrelativistic
limit (see Appendix \ref{sec:non-relativistic-limit}). In order to
recover the momentum dependence, one can express $\overline{\rho}_{00}^{\phi}$
in terms of lab-frame fields. The transformation of the fields between
the lab and rest frame reads
\begin{eqnarray}
{\bf B}_{\phi}^{\prime} & = & \gamma{\bf B}_{\phi}-\gamma{\bf v}\times{\bf E}_{\phi}+(1-\gamma)\frac{{\bf v}\cdot{\bf B}_{\phi}}{v^{2}}{\bf v},\nonumber \\
{\bf E}_{\phi}^{\prime} & = & \gamma{\bf E}_{\phi}+\gamma{\bf v}\times{\bf B}_{\phi}+(1-\gamma)\frac{{\bf v}\cdot{\bf E}_{\phi}}{v^{2}}{\bf v},\nonumber \\
\boldsymbol{\omega}^{\prime} & = & \gamma\boldsymbol{\omega}-\gamma{\bf v}\times\boldsymbol{\varepsilon}+(1-\gamma)\frac{{\bf v}\cdot\boldsymbol{\omega}}{v^{2}}{\bf v},\nonumber \\
\boldsymbol{\varepsilon}^{\prime} & = & \gamma\boldsymbol{\varepsilon}+\gamma{\bf v}\times\boldsymbol{\omega}+(1-\gamma)\frac{{\bf v}\cdot\boldsymbol{\varepsilon}}{v^{2}}{\bf v},
\end{eqnarray}
where $\gamma=E_{{\bf p}}^{\phi}/m_{\phi}$ is the Lorentz factor
and ${\bf v}=\mathbf{p}/E_{{\bf p}}^{\phi}$ is the velocity of the
$\phi$ meson. Taking the $y$-direction as the spin quantization
direction, $\boldsymbol{\epsilon}_{0}=(0,1,0)$, we obtain $\overline{\rho}_{00}^{\phi}$
in terms of the fields in the lab frame 
\begin{eqnarray}
\overline{\rho}_{00}^{\phi}(x,{\bf p}) & \approx & \frac{1}{3}+\frac{1}{3}\sum_{i=1,2,3}I_{B,i}({\bf p})\frac{1}{m_{\phi}^{2}}\left[\boldsymbol{\omega}_{i}^{2}-\frac{4g_{\phi}^{2}}{m_{\phi}^{2}T_{\mathrm{eff}}^{2}}(\mathbf{B}_{i}^{\phi})^{2}\right]\nonumber \\
 &  & +\frac{1}{3}\sum_{i=1,2,3}I_{E,i}({\bf p})\frac{1}{m_{\phi}^{2}}\left[\boldsymbol{\varepsilon}_{i}^{2}-\frac{4g_{\phi}^{2}}{m_{\phi}^{2}T_{\mathrm{eff}}^{2}}(\mathbf{E}_{i}^{\phi})^{2}\right],\label{eq:rho00-eb}
\end{eqnarray}
where the coefficients are given by 
\begin{eqnarray}
I_{B,x}({\bf p}) & = & C_{1}\left[(E_{{\bf p}}^{\phi})^{2}-\left(1+\frac{3p_{y}^{2}}{(m_{\phi}+E_{{\bf p}}^{\phi})^{2}}\right)p_{x}^{2}\right]+C_{2}(p_{y}^{2}-2p_{z}^{2}),\nonumber \\
I_{E,x}({\bf p}) & = & C_{1}(p_{y}^{2}-2p_{z}^{2})+C_{2}\left[(E_{{\bf p}}^{\phi})^{2}-\left(1+\frac{3p_{y}^{2}}{(m_{\phi}+E_{{\bf p}}^{\phi})^{2}}\right)p_{x}^{2}\right],\nonumber \\
I_{B,y}({\bf p}) & = & C_{1}\left[6\frac{E_{{\bf p}}^{\phi}}{m_{\phi}+E_{{\bf p}}^{\phi}}p_{y}^{2}-2(E_{{\bf p}}^{\phi})^{2}-p_{y}^{2}-\frac{3p_{y}^{4}}{(m_{\phi}+E_{{\bf p}}^{\phi})^{2}}\right]+C_{2}(p_{x}^{2}+p_{z}^{2}),\nonumber \\
I_{E,y}({\bf p}) & = & C_{1}(p_{x}^{2}+p_{z}^{2})+C_{2}\left[6\frac{E_{{\bf p}}^{\phi}}{m_{\phi}+E_{{\bf p}}^{\phi}}p_{y}^{2}-2(E_{{\bf p}}^{\phi})^{2}-p_{y}^{2}-\frac{3p_{y}^{4}}{(m_{\phi}+E_{{\bf p}}^{\phi})^{2}}\right],\nonumber \\
I_{B,z}({\bf p}) & = & C_{1}\left[(E_{{\bf p}}^{\phi})^{2}-\left(1+\frac{3p_{y}^{2}}{(m_{\phi}+E_{{\bf p}}^{\phi})^{2}}\right)p_{z}^{2}\right]+C_{2}(p_{y}^{2}-2p_{x}^{2}),\nonumber \\
I_{E,z}({\bf p}) & = & C_{1}(p_{y}^{2}-2p_{x}^{2})+C_{2}\left[(E_{{\bf p}}^{\phi})^{2}-\left(1+\frac{3p_{y}^{2}}{(m_{\phi}+E_{{\bf p}}^{\phi})^{2}}\right)p_{z}^{2}\right].
\end{eqnarray}
The result in Eq. (\ref{eq:rho00-eb}) is remarkable in its factorization
form: the momentum functions are separated from space-time functions.
This has an advantage that the momentum functions can be determined
by experimental data on momentum spectra while unknown space-time
functions can be extracted from data on $\overline{\rho}_{00}^{\phi}$.

One can take an average of $\overline{\rho}_{00}^{\phi}(x,{\bf p})$
over the local space-time volume in which the vector meson is formed
as 
\begin{eqnarray}
\left\langle \overline{\rho}_{00}^{\phi}(x,{\bf p})\right\rangle _{x} & \approx & \frac{1}{3}+\frac{1}{3}\sum_{i=1,2,3}I_{B,i}({\bf p})\frac{1}{m_{\phi}^{2}}\left[\left\langle \boldsymbol{\omega}_{i}^{2}\right\rangle -\frac{4g_{\phi}^{2}}{m_{\phi}^{2}T_{\mathrm{eff}}^{2}}\left\langle (\mathbf{B}_{i}^{\phi})^{2}\right\rangle \right]\nonumber \\
 &  & +\frac{1}{3}\sum_{i=1,2,3}I_{E,i}({\bf p})\frac{1}{m_{\phi}^{2}}\left[\left\langle \boldsymbol{\varepsilon}_{i}^{2}\right\rangle -\frac{4g_{\phi}^{2}}{m_{\phi}^{2}T_{\mathrm{eff}}^{2}}\left\langle (\mathbf{E}_{i}^{\phi})^{2}\right\rangle \right].\label{eq:rho00-x}
\end{eqnarray}
These averaged field squares can play as parameters and be determined
by comparing $\left\langle \overline{\rho}_{00}^{\phi}(x,{\bf p})\right\rangle _{x}$
with the data of $\rho_{00}^{\phi}$ as functions of transverse momenta.
One can further take a momentum average of $\left\langle \overline{\rho}_{00}^{\phi}(x,{\bf p})\right\rangle _{x}$
and compare with the data as functions of collision energies,
\begin{eqnarray}
\left\langle \overline{\rho}_{00}^{\phi}(x,{\bf p})\right\rangle _{x,\mathbf{p}} & \approx & \frac{1}{3}+\frac{1}{3}\sum_{i=1,2,3}\left\langle I_{B,i}({\bf p})\right\rangle \frac{1}{m_{\phi}^{2}}\left[\left\langle \boldsymbol{\omega}_{i}^{2}\right\rangle -\frac{4g_{\phi}^{2}}{m_{\phi}^{2}T_{\mathrm{eff}}^{2}}\left\langle (\mathbf{B}_{i}^{\phi})^{2}\right\rangle \right]\nonumber \\
 &  & +\frac{1}{3}\sum_{i=1,2,3}\left\langle I_{E,i}({\bf p})\right\rangle \frac{1}{m_{\phi}^{2}}\left[\left\langle \boldsymbol{\varepsilon}_{i}^{2}\right\rangle -\frac{4g_{\phi}^{2}}{m_{\phi}^{2}T_{\mathrm{eff}}^{2}}\left\langle (\mathbf{E}_{i}^{\phi})^{2}\right\rangle \right],\label{eq:rho00-xp}
\end{eqnarray}
where the momentum average is defined as 
\begin{equation}
\left\langle O(\mathbf{p})\right\rangle =\frac{\int d^{3}\mathbf{p}O(\mathbf{p})f_{\phi}(\mathbf{p})}{\int d^{3}\mathbf{p}f_{\phi}(\mathbf{p})},
\end{equation}
with $f_{\phi}(\mathbf{p})$ being the momentum distribution of the
$\phi$ meson which can be determined by experimental data. The theoretical
results for $\rho_{00}^{\phi}$ as functions of transverse momenta,
collision energies and centralities are presented in Ref. \citep{Sheng:2022wsy},
which are in a good agreement with recent STAR data \citep{STAR:2022fan}.

\section{Discussions and conclusions}

\label{sec:discussions}In this section we will discuss about the
main results as well as approximations or assumptions that have been
made in this paper.

The Lagrangian (\ref{eq:lag-vector}) is for real vector fields since
we are concerned about the charge or flavor neutral particles such
as quarkonia made of a quark and its antiquark. To describe those
particles that carry net charge or flavor, we have to consider complex
vector fields. The generalization of the formalism to complex vector
fields is straightforward.

The vector fields that polarize $s$ and $\overline{s}$ are assumed
to be the $\phi$ fields, the effective (color singlet) modes of the
strong force that carry vacuum quantum number. As an input to the
general formula (\ref{eq:spin-density-sz1-sz2}) we assume that $P_{q}^{\mu}$
and $P_{\overline{q}}^{\mu}$ have the linear form in Eq. (\ref{eq:polarization-ssbar})
in the vorticity and $\phi$ fields. The coupling between the spin
and fluid velocity field is assumed to be through the vorticity. One
can also introduce other coupling forms such as spin-shear couplings
\citep{Liu:2021uhn,Becattini:2021suc,Fu:2021pok,Becattini:2021iol}.
The spin coupling to the $\phi$ field is assumed to have a covariant
form $\sim\epsilon^{\mu\nu\alpha\beta}F_{\alpha\beta}^{\phi}p_{\nu}$.
This is one of our main assumptions. Of course one can use other forms
of spin-field couplings or add more terms to Eq. (\ref{eq:polarization-ssbar}).
An alternative choice is to use the coupling of the spin and gluon
field as in the nonrelativistic quantum chromodynamics (NRQCD) \citep{Bodwin:1994jh,Braaten:1996ix}.
But the Hamiltonian of NRQCD is not covariant at all and may be different
from the covariant from $\sim\epsilon^{\mu\nu\alpha\beta}F_{\alpha\beta}^{c}p_{\nu}$
where $F_{\alpha\beta}^{c}$ is the gluon field with adjoint color
$c$. In this case the final result may be different from the result
in this paper.

If we use gluon fields that are coupled to the spin, the MVSD for
the quark and antiquark in (\ref{eq:f-rs-pol}) and the polarization
vectors in (\ref{eq:polarization-ssbar}) have to be modified to include
color indices. The effect to $\rho_{\lambda_{1}\lambda_{2}}^{V}$
in (\ref{eq:spin-density-sz1-sz2}) is to replace $P_{q}^{\mu}\rightarrow P_{q}^{\mu,c}$
and $P_{\overline{q}}^{\mu}\rightarrow P_{\overline{q}}^{\mu,c}$
with $c$ denoting the adjoint color of the gluon and to sum over
the gluon color $c$ to accommodate the color singlet requirement.
Correspondingly, in Eq. (\ref{eq:polarization-ssbar}), we replace
$P_{s}^{\mu}\rightarrow P_{s}^{\mu,c}$, $P_{\overline{s}}^{\mu}\rightarrow P_{\overline{s}}^{\mu,c}$,
and $F_{\rho\sigma}^{\phi}\rightarrow F_{\rho\sigma}^{c}\lambda^{c}/2$,
where $\lambda^{c}$ are Gell-mann matrices. The effect to $\rho_{00}^{\phi}(F_{\phi}^{2})$
in (\ref{eq:rho-f2}) is to replace $\widetilde{F}_{\rho\xi}^{\phi}(x)\widetilde{F}_{\sigma\gamma}^{\phi}(x)\rightarrow\sum_{c}\widetilde{F}_{\rho\xi}^{c}(x)\widetilde{F}_{\sigma\gamma}^{c}(x)$
up to a color factor, which results in the replacements $(\mathbf{E}_{i}^{\phi})^{2}\rightarrow\sum_{c}(\mathbf{E}_{i}^{c})^{2}$
and $(\mathbf{B}_{i}^{\phi})^{2}\rightarrow\sum_{c}(\mathbf{B}_{i}^{c})^{2}$
in the final results in Eqs. (\ref{eq:rho00-eb})-(\ref{eq:rho00-xp}),
where $\mathbf{E}^{c}$ and $\mathbf{B}^{c}$ are the chromoelectric
and chromomagnetic fields respectively. 

In evaluating the integrals in momentum moments in the rest frame
of the vector meson, we assume a simple form for the fluid four-velocity
$u^{\prime\mu}=(1,\mathbf{0})$ so that the quark and antiquark distributions
depend only on energies. Then we obtain the simple form of $\overline{\rho}_{00}^{\phi}(x,{\bf p})$
in Eq. (\ref{eq:av-rho00-square}) with $C_{1}$ and $C_{2}$ depending
only on masses as shown in (\ref{eq:constant-mass}). In general the
fluid four-velocity has also a spatial component or three-velocity,
in this case $\overline{\rho}_{00}^{\phi}(x,{\bf p})$ should have
much more complicated form than Eq. (\ref{eq:av-rho00-square}) where
the coefficients also depend on the three-velocity of the fluid in
a more sophisticated way.

According to the chiral quark model in Ref. \citep{Manohar:1983md}
the local averaged field squares $\left\langle (\mathbf{B}_{i}^{\phi})^{2}\right\rangle $
and $\left\langle (\mathbf{E}_{i}^{\phi})^{2}\right\rangle $ are
related to the fields of psuedo-Goldstone bosons. They are also related
to gluon fluctuation of instantons \citep{Shuryak:1981ff,Shuryak:1982dp}
according to the quark model based on instanton vacuum \citep{Diakonov:1985eg}.
If quarks and antiquarks are polarized by gluon fields, the local
averaged field squares are related to the gluon condensate which contributes
to the trace anomaly of the energy momentum tensor. Therefore the
local averaged field squares are in connection with fundamental properties
of the QCD vacuum which play an important role in hadron structures
\citep{Shifman:1978bx,Shifman:1978by}.

In summary, a relativistic theory for the spin density matrix of vector
mesons is constructed based on Kadanoff-Baym (KB) equations from which
the spin Boltzmann equations are derived. With the spin Boltzmann
equations we formulate the spin density matrix element $\rho_{00}$
for $\phi$ mesons. The dominant contributions to $\rho_{00}^{\phi}$
at lower energies are assumed to come from the $\phi$ field, a kind
of the strong force field that can polarize the strange quark and
antiquark in the same way as the electromagnetic field. The key observation
is that there is correlation inside the $\phi$ meson wave function
between the $\phi$ field that polarizes the strange quark and that
polarizes the strange antiquark. This is reflected by the fact that
the contributions to $\rho_{00}^{\phi}$ are all in squares of the
fields which are nonvanishing even if the fields may strongly fluctuate.
Then the fluctuations of strong force fields can be extracted from
$\rho_{00}$ of quarkonium vector mesons as links to fundamental properties
of QCD.

\section*{Acknowledgement}

The authors thank C.D. Roberts for providing us with the Bethe-Salpeter
wave function of the $\phi$ meson. The authors thank X. G. Huang,
J. F. Liao, S. Pu, A. H. Tang, D. L. Yang and Y. Yin for helpful discussion.
This work is supported in part by the National Natural Science Foundation
of China (NSFC) under Grants No. 12135011, 11890713 (a subgrant of
11890710), the Strategic Priority Research Program of the Chinese
Academy of Sciences (CAS) under Grant No. XDB34030102, and by the
Director, Office of Energy Research, Office of High Energy and Nuclear
Physics, Division of Nuclear Physics, of the U.S. Department of Energy
under Contract No. DE-AC02-05CH11231.

\appendix

\section{Collision terms for coalescence and dissociation}

\label{sec:coalescence}In this appendix, we will derive the collision
term for the coalescence process of the quark and antiquark into the
vector meson corresponding to $I_{-++}$ in Eq. (\ref{eq:boltzmann-all}).

The explicit form of $I_{-++}$ is 
\begin{eqnarray}
I_{-++} & = & \sum_{r_{1},s_{1},r_{2},s_{2},\lambda_{1}^{\prime},\lambda_{2}^{\prime}}\left\{ \epsilon^{\alpha}\left(\lambda_{1}^{\prime},{\bf p}\right)\epsilon^{\nu\ast}\left(\lambda_{2}^{\prime},{\bf p}\right)\right.\nonumber \\
 &  & \times\text{Tr}\left[\Gamma_{\alpha}v(s_{1},-{\bf p}^{\prime})\overline{v}(r_{1},-{\bf p}^{\prime})\Gamma^{\mu}u(r_{2},\mathbf{p}+\mathbf{p}^{\prime})\overline{u}(s_{2},\mathbf{p}+\mathbf{p}^{\prime})\right]\nonumber \\
 &  & +\epsilon^{\mu}\left(\lambda_{1}^{\prime},{\bf p}\right)\epsilon_{\alpha}^{\ast}\left(\lambda_{2}^{\prime},{\bf p}\right)\nonumber \\
 &  & \times\left.\text{Tr}\left[\Gamma^{\nu}v(s_{1},-{\bf p}^{\prime})\overline{v}(r_{1},-{\bf p}^{\prime})\Gamma^{\alpha}u(r_{2},\mathbf{p}+\mathbf{p}^{\prime})\overline{u}(s_{2},\mathbf{p}+\mathbf{p}^{\prime})\right]\right\} \nonumber \\
 &  & \times\left\{ \left[\delta_{r_{1}s_{1}}-f_{r_{1}s_{1}}^{(-)}(x,-{\bf p}^{\prime})\right]\left[\delta_{r_{2}s_{2}}-f_{r_{2}s_{2}}^{(+)}(x,\mathbf{p}+\mathbf{p}^{\prime})\right]f_{\lambda_{1}^{\prime}\lambda_{2}^{\prime}}(x,\mathbf{p})\right.\nonumber \\
 &  & \left.-f_{r_{1}s_{1}}^{(-)}(x,-{\bf p}^{\prime})f_{r_{2}s_{2}}^{(+)}(x,\mathbf{p}+\mathbf{p}^{\prime})\left[\delta_{\lambda_{1}^{\prime}\lambda_{2}^{\prime}}+f_{\lambda_{1}^{\prime}\lambda_{2}^{\prime}}(x,\mathbf{p})\right]\right\} .
\end{eqnarray}
The corresponding collision term reads 
\begin{eqnarray}
C_{\text{coalescence}}^{\mu\nu} & = & \frac{1}{4(2\pi\hbar)}\sum_{r_{1},s_{1},r_{2},s_{2},\lambda_{1}^{\prime},\lambda_{2}^{\prime}}\int d^{3}\mathbf{p}^{\prime}\frac{1}{8E_{p^{\prime}}^{\overline{q}}E_{{\bf p}-{\bf p}^{\prime}}^{q}E_{p}^{V}}\delta\left(E_{p}^{V}-E_{p^{\prime}}^{\overline{q}}-E_{{\bf p}-{\bf p}^{\prime}}^{q}\right)\delta(p^{0}-E_{p}^{V})\nonumber \\
 &  & \times\left\{ \epsilon^{\alpha}\left(\lambda_{1}^{\prime},{\bf p}\right)\epsilon^{\nu\ast}\left(\lambda_{2}^{\prime},{\bf p}\right)\text{Tr}\left[\Gamma_{\alpha}v(s_{1},{\bf p}^{\prime})\overline{v}(r_{1},{\bf p}^{\prime})\Gamma^{\mu}u(r_{2},\mathbf{p}-{\bf p}^{\prime})\overline{u}(s_{2},\mathbf{p}-{\bf p}^{\prime})\right]\right.\nonumber \\
 &  & +\left.\epsilon^{\mu}\left(\lambda_{1}^{\prime},{\bf p}\right)\epsilon_{\alpha}^{\ast}\left(\lambda_{2}^{\prime},{\bf p}\right)\text{Tr}\left[\Gamma^{\nu}v(s_{1},{\bf p}^{\prime})\overline{v}(r_{1},{\bf p}^{\prime})\Gamma^{\alpha}u(r_{2},\mathbf{p}-{\bf p}^{\prime})\overline{u}(s_{2},\mathbf{p}-{\bf p}^{\prime})\right]\right\} \nonumber \\
 &  & \times\left\{ \left[\delta_{r_{1}s_{1}}-f_{r_{1}s_{1}}^{(-)}(x,{\bf p}^{\prime})\right]\left[\delta_{r_{2}s_{2}}-f_{r_{2}s_{2}}^{(+)}(x,\mathbf{p}-\mathbf{p}^{\prime})\right]f_{\lambda_{1}^{\prime}\lambda_{2}^{\prime}}(x,\mathbf{p})\right.\nonumber \\
 &  & \left.-f_{r_{1}s_{1}}^{(-)}(x,{\bf p}^{\prime})f_{r_{2}s_{2}}^{(+)}(x,\mathbf{p}-\mathbf{p}^{\prime})\left[\delta_{\lambda_{1}^{\prime}\lambda_{2}^{\prime}}+f_{\lambda_{1}^{\prime}\lambda_{2}^{\prime}}(x,\mathbf{p})\right]\right\} ,\label{eq:coalescence-1}
\end{eqnarray}
where we have changed the sign of the antiquark's three-momentum as
${\bf p}^{\prime}\rightarrow-{\bf p}^{\prime}$ in the integral, used
Eq. (\ref{eq:polar-vector-rel}) and the relation 
\begin{eqnarray}
 &  & \delta(p^{\prime2}-m_{q}^{2})\delta\left[(p+p^{\prime})^{2}-m_{q}^{2}\right]\delta(p^{2}-m_{V}^{2})\theta(-p_{0}^{\prime})\theta\left(p_{0}+p_{0}^{\prime}\right)\theta(p_{0})\nonumber \\
 & = & \frac{1}{8E_{p^{\prime}}^{\overline{q}}E_{{\bf p}+{\bf p}^{\prime}}^{q}E_{p}^{V}}\delta(p_{0}^{\prime}+E_{p^{\prime}}^{\overline{q}})\delta\left(p_{0}+p_{0}^{\prime}-E_{{\bf p}+{\bf p}^{\prime}}^{q}\right)\delta(p_{0}-E_{p}^{V})\nonumber \\
 & = & \frac{1}{8E_{p^{\prime}}^{\overline{q}}E_{{\bf p}+{\bf p}^{\prime}}^{q}E_{p}^{V}}\delta(p_{0}^{\prime}+E_{p^{\prime}}^{\overline{q}})\delta\left(E_{p}^{V}-E_{p^{\prime}}^{\overline{q}}-E_{{\bf p}+{\bf p}^{\prime}}^{q}\right)\delta(p_{0}-E_{p}^{V}).
\end{eqnarray}

From (\ref{eq:wigner-func-vm}), the particle sector of $p\cdot\partial_{x}G^{<,\mu\nu}(x,p)$
in the left-hand-side of Eq. (\ref{eq:boltzmann-all}) becomes 
\begin{eqnarray}
p\cdot\partial_{x}G^{<,\mu\nu}(x,p) & = & 2\pi\hbar\frac{1}{2E_{p}^{V}}\delta(p_{0}-E_{p}^{V})\nonumber \\
 &  & \times\sum_{\lambda_{1}^{\prime},\lambda_{2}^{\prime}}\epsilon^{\mu}\left(\lambda_{1}^{\prime},{\bf p}\right)\epsilon^{\nu\ast}\left(\lambda_{2}^{\prime},{\bf p}\right)p\cdot\partial_{x}f_{\lambda_{1}^{\prime}\lambda_{2}^{\prime}}(x,\mathbf{p}).\label{eq:kinetic-term-vm}
\end{eqnarray}
Using Eqs. (\ref{eq:coalescence-1}) and (\ref{eq:kinetic-term-vm})
into Eq. (\ref{eq:boltzmann-all}), taking a contraction of the resulting
equation with $\epsilon_{\mu}^{\ast}(\lambda_{1},{\bf p})$ and $\epsilon_{\nu}(\lambda_{2},{\bf p})$,
and using the first identity in (\ref{eq:polar-vector-rel}), we obtain
\begin{eqnarray}
 &  & p\cdot\partial_{x}f_{\lambda_{1}\lambda_{2}}(x,\mathbf{p})\nonumber \\
 & = & \frac{1}{16}\sum_{r_{1},s_{1},r_{2},s_{2},\lambda_{1}^{\prime},\lambda_{2}^{\prime}}\int\frac{d^{3}\mathbf{p}^{\prime}}{(2\pi\hbar)^{3}}\frac{1}{E_{p^{\prime}}^{\overline{q}}E_{{\bf p}-{\bf p}^{\prime}}^{q}}2\pi\hbar\delta\left(E_{p}^{V}-E_{p^{\prime}}^{\overline{q}}-E_{{\bf p}-{\bf p}^{\prime}}^{q}\right)\nonumber \\
 &  & \times\left\{ \delta_{\lambda_{2}\lambda_{2}^{\prime}}\epsilon_{\mu}^{\ast}(\lambda_{1},{\bf p})\epsilon^{\alpha}\left(\lambda_{1}^{\prime},{\bf p}\right)\text{Tr}\left[\Gamma_{\alpha}v(s_{1},{\bf p}^{\prime})\overline{v}(r_{1},{\bf p}^{\prime})\Gamma^{\mu}u(r_{2},\mathbf{p}-{\bf p}^{\prime})\overline{u}(s_{2},\mathbf{p}-{\bf p}^{\prime})\right]\right.\nonumber \\
 &  & +\left.\delta_{\lambda_{1}\lambda_{1}^{\prime}}\epsilon_{\nu}(\lambda_{2},{\bf p})\epsilon_{\alpha}^{\ast}\left(\lambda_{2}^{\prime},{\bf p}\right)\text{Tr}\left[\Gamma^{\nu}v(s_{1},{\bf p}^{\prime})\overline{v}(r_{1},{\bf p}^{\prime})\Gamma^{\alpha}u(r_{2},\mathbf{p}-{\bf p}^{\prime})\overline{u}(s_{2},\mathbf{p}-{\bf p}^{\prime})\right]\right\} \nonumber \\
 &  & \times\left\{ f_{r_{1}s_{1}}^{(-)}(x,{\bf p}^{\prime})f_{r_{2}s_{2}}^{(+)}(x,\mathbf{p}-\mathbf{p}^{\prime})\left[\delta_{\lambda_{1}^{\prime}\lambda_{2}^{\prime}}+f_{\lambda_{1}^{\prime}\lambda_{2}^{\prime}}(x,\mathbf{p})\right]\right.\nonumber \\
 &  & \left.-\left[\delta_{r_{1}s_{1}}-f_{r_{1}s_{1}}^{(-)}(x,{\bf p}^{\prime})\right]\left[\delta_{r_{2}s_{2}}-f_{r_{2}s_{2}}^{(+)}(x,\mathbf{p}-\mathbf{p}^{\prime})\right]f_{\lambda_{1}^{\prime}\lambda_{2}^{\prime}}(x,\mathbf{p})\right\} ,\label{eq:boltz-mvsd}
\end{eqnarray}
which reproduces Eq. (\ref{eq:boltz-mvsd-1}). Note that the terms
proportional to $p^{\mu}$ and $p^{\nu}$ in the left-hand-side of
Eq. (\ref{eq:boltzmann-all}) do not contribute since their contraction
with $\epsilon_{\mu}^{\ast}(\lambda_{1},{\bf p})$ and $\epsilon_{\nu}(\lambda_{2},{\bf p})$
is vanishing.  

We consider the coalescence process in heavy ion collisions in which
the MVSDs of quarks, antiquarks and vector mesons are assumed to be
much smaller than unity. So the term with $\delta_{\lambda_{1}\lambda_{2}}$
dominates the gain term which can be simplified as 
\begin{eqnarray}
\mathrm{gain} & \approx & \frac{1}{8(2\pi\hbar)^{2}}\sum_{r_{1},s_{1},r_{2},s_{2}}\int d^{3}\mathbf{p}^{\prime}\frac{1}{E_{p^{\prime}}^{\overline{q}}E_{{\bf p}-{\bf p}^{\prime}}^{q}}\delta\left(E_{p}^{V}-E_{p^{\prime}}^{\overline{q}}-E_{{\bf p}-{\bf p}^{\prime}}^{q}\right)\nonumber \\
 &  & \times\epsilon_{\mu}^{\ast}\left(\lambda_{1},{\bf p}\right)\epsilon_{\alpha}\left(\lambda_{2},{\bf p}\right)\text{Tr}\left[\Gamma^{\alpha}v(s_{1},{\bf p}^{\prime})\overline{v}(r_{1},{\bf p}^{\prime})\Gamma^{\mu}u(r_{2},\mathbf{p}-{\bf p}^{\prime})\overline{u}(s_{2},\mathbf{p}-{\bf p}^{\prime})\right]\nonumber \\
 &  & \times f_{r_{1}s_{1}}^{(-)}(x,{\bf p}^{\prime})f_{r_{2}s_{2}}^{(+)}(x,\mathbf{p}-\mathbf{p}^{\prime}),\label{eq:gain}
\end{eqnarray}
which gives Eq. (\ref{eq:coal}). The loss term can be simplified
as 
\begin{eqnarray}
\mathrm{loss} & \approx & -\frac{1}{16(2\pi\hbar)^{2}}\sum_{r_{1},s_{1},r_{2},s_{2}}\int d^{3}\mathbf{p}^{\prime}\frac{1}{E_{p^{\prime}}^{\overline{q}}E_{{\bf p}-{\bf p}^{\prime}}^{q}}\delta\left(E_{p}^{V}-E_{p^{\prime}}^{\overline{q}}-E_{{\bf p}-{\bf p}^{\prime}}^{q}\right)\nonumber \\
 &  & \times\left\{ \delta_{\lambda_{2}\lambda_{2}^{\prime}}\epsilon_{\mu}^{\ast}\left(\lambda_{1},{\bf p}\right)\epsilon^{\alpha}\left(\lambda_{1}^{\prime},{\bf p}\right)\text{Tr}\left[\Gamma_{\alpha}v(s_{1},{\bf p}^{\prime})\overline{v}(r_{1},{\bf p}^{\prime})\Gamma^{\mu}u(r_{2},\mathbf{p}-{\bf p}^{\prime})\overline{u}(s_{2},\mathbf{p}-{\bf p}^{\prime})\right]\right.\nonumber \\
 &  & +\left.\delta_{\lambda_{1}\lambda_{1}^{\prime}}\epsilon_{\nu}\left(\lambda_{2},{\bf p}\right)\epsilon_{\alpha}^{\ast}\left(\lambda_{2}^{\prime},{\bf p}\right)\text{Tr}\left[\Gamma^{\nu}v(s_{1},{\bf p}^{\prime})\overline{v}(r_{1},{\bf p}^{\prime})\Gamma^{\alpha}u(r_{2},\mathbf{p}-{\bf p}^{\prime})\overline{u}(s_{2},\mathbf{p}-{\bf p}^{\prime})\right]\right\} \nonumber \\
 &  & \times\delta_{r_{1}s_{1}}\delta_{r_{2}s_{2}}f_{\lambda_{1}^{\prime}\lambda_{2}^{\prime}}(x,\mathbf{p}),\nonumber \\
 & = & -\frac{1}{16(2\pi\hbar)^{2}}\int d^{3}\mathbf{p}^{\prime}\frac{1}{E_{p^{\prime}}^{\overline{q}}E_{{\bf p}-{\bf p}^{\prime}}^{q}}\delta\left(E_{p}^{V}-E_{p^{\prime}}^{\overline{q}}-E_{{\bf p}-{\bf p}^{\prime}}^{q}\right)\nonumber \\
 &  & \times\left\{ \sum_{\lambda_{1}^{\prime}}f_{\lambda_{1}^{\prime}\lambda_{2}}(x,\mathbf{p})\epsilon_{\mu}^{\ast}\left(\lambda_{1},{\bf p}\right)\epsilon^{\alpha}\left(\lambda_{1}^{\prime},{\bf p}\right)\text{Tr}\left[\Gamma_{\alpha}\left(p^{\prime}\cdot\gamma-m_{\overline{q}}\right)\Gamma^{\mu}\left((p-p^{\prime})\cdot\gamma+m_{q}\right)\right]\right.\nonumber \\
 &  & +\left.\sum_{\lambda_{2}^{\prime}}f_{\lambda_{1}\lambda_{2}^{\prime}}(x,\mathbf{p})\epsilon_{\nu}\left(\lambda_{2},{\bf p}\right)\epsilon_{\alpha}^{\ast}\left(\lambda_{2}^{\prime},{\bf p}\right)\text{Tr}\left[\Gamma^{\nu}\left(p^{\prime}\cdot\gamma-m_{\overline{q}}\right)\Gamma^{\alpha}\left((p-p^{\prime})\cdot\gamma+m_{q}\right)\right]\right\} \nonumber \\
 & = & -\frac{1}{16(2\pi\hbar)^{2}}\int d^{3}\mathbf{p}^{\prime}\frac{1}{E_{p^{\prime}}^{\overline{q}}E_{{\bf p}-{\bf p}^{\prime}}^{q}}\delta\left(E_{p}^{V}-E_{p^{\prime}}^{\overline{q}}-E_{{\bf p}-{\bf p}^{\prime}}^{q}\right)\nonumber \\
 &  & \times\left[\sum_{\lambda_{1}^{\prime}}f_{\lambda_{1}^{\prime}\lambda_{2}}(x,\mathbf{p})\epsilon_{\mu}^{\ast}\left(\lambda_{1},{\bf p}\right)\epsilon_{\alpha}\left(\lambda_{1}^{\prime},{\bf p}\right)+\sum_{\lambda_{2}^{\prime}}f_{\lambda_{1}\lambda_{2}^{\prime}}(x,\mathbf{p})\epsilon_{\mu}^{\ast}\left(\lambda_{2}^{\prime},{\bf p}\right)\epsilon_{\alpha}\left(\lambda_{2},{\bf p}\right)\right]\nonumber \\
 &  & \times\text{Tr}\left\{ \Gamma^{\alpha}\left(p^{\prime}\cdot\gamma-m_{\overline{q}}\right)\Gamma^{\mu}\left[(p-p^{\prime})\cdot\gamma+m_{q}\right]\right\} ,\label{eq:loss}
\end{eqnarray}
where we have neglected $f_{r_{1}s_{1}}^{(-)}$ and $f_{r_{2}s_{2}}^{(+)}$
relative to $\delta_{r_{1}s_{1}}$ and $\delta_{r_{2}s_{2}}$ respectively.
After completing the integral over $\mathbf{p}^{\prime}$ in the vector
meson's rest frame, one can prove 
\begin{eqnarray}
 &  & \epsilon_{\mu}^{\ast}\left(\lambda_{1},{\bf p}\right)\epsilon_{\alpha}\left(\lambda_{1}^{\prime},{\bf p}\right)\int d^{3}\mathbf{p}^{\prime}\frac{1}{4E_{p^{\prime}}^{\overline{q}}E_{{\bf p}-{\bf p}^{\prime}}^{q}}\delta\left(E_{p}^{V}-E_{p^{\prime}}^{\overline{q}}-E_{{\bf p}-{\bf p}^{\prime}}^{q}\right)\nonumber \\
 &  & \times\text{Tr}\left\{ \Gamma^{\alpha}\left(p^{\prime}\cdot\gamma-m_{\overline{q}}\right)\Gamma^{\mu}\left[(p-p^{\prime})\cdot\gamma+m_{q}\right]\right\} \propto\delta_{\lambda_{1}\lambda_{1}^{\prime}},
\end{eqnarray}
so we can replace 
\begin{eqnarray}
 &  & \sum_{\lambda_{1}^{\prime}}f_{\lambda_{1}^{\prime}\lambda_{2}}(x,\mathbf{p})\epsilon_{\mu}^{\ast}\left(\lambda_{1},{\bf p}\right)\epsilon_{\alpha}\left(\lambda_{1}^{\prime},{\bf p}\right)+\sum_{\lambda_{2}^{\prime}}f_{\lambda_{1}^{\prime}\lambda_{2}}(x,\mathbf{p})\epsilon_{\mu}^{\ast}\left(\lambda_{2},{\bf p}\right)\epsilon_{\alpha}(\lambda_{2}^{\prime},{\bf p})\nonumber \\
 & \rightarrow & -\frac{2}{3}f_{\lambda_{1}\lambda_{2}}(x,\mathbf{p})\left(g_{\mu\alpha}-\frac{p_{\mu}p_{\alpha}}{m_{V}^{2}}\right),
\end{eqnarray}
then the loss term becomes 
\begin{eqnarray}
\mathrm{loss} & \approx & \frac{1}{12(2\pi\hbar)^{2}}f_{\lambda_{1}\lambda_{2}}(x,\mathbf{p})\left(g_{\mu\alpha}-\frac{p_{\mu}p_{\alpha}}{m_{V}^{2}}\right)\int d^{3}\mathbf{p}^{\prime}\frac{1}{E_{p^{\prime}}^{\overline{q}}E_{{\bf p}-{\bf p}^{\prime}}^{q}}\delta\left(E_{p}^{V}-E_{p^{\prime}}^{\overline{q}}-E_{{\bf p}-{\bf p}^{\prime}}^{q}\right)\nonumber \\
 &  & \times\text{Tr}\left\{ \Gamma^{\alpha}\left(p^{\prime}\cdot\gamma-m_{\overline{q}}\right)\Gamma^{\mu}\left[(p-p^{\prime})\cdot\gamma+m_{q}\right]\right\} .\label{eq:loss-1}
\end{eqnarray}
This gives Eq. (\ref{eq:dis-loss}).

% done. 2021.03.08. 14:00

\section{Collision kernel}

\label{sec:collision-kernel}The spin density matrix element for vector
mesons is given by Eq. (\ref{eq:spin-density-sz1-sz2}). In this appendix
we evaluate the collision kernel in Eq. (\ref{eq:spin-density-sz1-sz2})
\begin{eqnarray}
I_{\lambda_{1}\lambda_{2}}({\bf p},{\bf p}^{\prime}) & = & I^{\alpha\beta}({\bf p},{\bf p}^{\prime})\epsilon_{\alpha}^{\ast}(\lambda_{1},{\bf p})\epsilon_{\beta}(\lambda_{2},{\bf p}),\label{eq:i-l1l2}
\end{eqnarray}
where $I^{\alpha\beta}({\bf p},{\bf p}^{\prime})$ is defined as 
\begin{eqnarray}
I^{\alpha\beta}({\bf p},{\bf p}^{\prime}) & \equiv & \text{Tr}\left[\Gamma^{\beta}v(s_{1},\mathbf{p}^{\prime})\overline{v}(r_{1},\mathbf{p}^{\prime})\Gamma^{\alpha}u(r_{2},\mathbf{p}-\mathbf{p}^{\prime})\overline{u}(s_{2},\mathbf{p}-\mathbf{p}^{\prime})\right]\nonumber \\
 &  & \times f_{r_{1}s_{1}}^{(\overline{q})}(x,\mathbf{p}^{\prime})f_{r_{2}s_{2}}^{(q)}(x,\mathbf{p}-\mathbf{p}^{\prime}).\label{eq:isz1-sz2}
\end{eqnarray}
Now we use following formula to simplify $I^{\alpha\beta}$. For quark
spinors of particles and antiparticles we have 
\begin{eqnarray}
u(r,\mathbf{p})\overline{u}(s,\mathbf{p}) & = & \frac{1}{2}\left(m_{q}+\gamma^{\mu}p_{\mu}\right)\delta_{rs}+\frac{1}{2}m_{q}\gamma^{5}\gamma^{\mu}n_{\mu}(\mathbf{n}_{sr},\mathbf{p},m_{q})\nonumber \\
 &  & -\frac{1}{4}\epsilon_{\mu\nu\alpha\beta}\sigma^{\mu\nu}p^{\alpha}n^{\beta}(\mathbf{n}_{sr},\mathbf{p},m_{q}),\nonumber \\
\sum_{r,s}u(r,\mathbf{p})\overline{u}(s,\mathbf{p})(\tau_{j})_{rs} & = & m_{q}\gamma^{5}\gamma^{\mu}n_{\mu}(\mathbf{n}_{j},\mathbf{p},m_{q})-\frac{1}{2}\epsilon_{\mu\nu\alpha\beta}\sigma^{\mu\nu}p^{\alpha}n^{\beta}(\mathbf{n}_{j},\mathbf{p},m_{q}),\nonumber \\
v(r,\mathbf{p})\overline{v}(s,\mathbf{p}) & = & \frac{1}{2}(-m_{\overline{q}}+p^{\mu}\gamma_{\mu})\delta_{rs}-\frac{1}{2}m_{\overline{q}}\gamma_{5}\gamma_{\mu}n^{\mu}(\mathbf{n}_{sr}^{*},\mathbf{p},m_{\overline{q}})\nonumber \\
 &  & -\frac{1}{4}\epsilon^{\mu\nu\alpha\beta}\sigma_{\mu\nu}p_{\alpha}n_{\beta}(\mathbf{n}_{sr}^{*},\mathbf{p},m_{\overline{q}}),\nonumber \\
\sum_{r,s}v(r,\mathbf{p})\overline{v}(s,\mathbf{p})(\tau_{j})_{sr} & = & -m_{\overline{q}}\gamma_{5}\gamma_{\mu}n^{\mu}(\mathbf{n}_{j},\mathbf{p},m_{\overline{q}})-\frac{1}{2}\epsilon^{\mu\nu\alpha\beta}\sigma_{\mu\nu}p_{\alpha}n_{\beta}(\mathbf{n}_{j},\mathbf{p},m_{\overline{q}}),
\end{eqnarray}
where we have used $\mathbf{n}_{sr}^{*}=\mathbf{n}_{rs}=\mathbf{n}_{i}(\tau_{i})_{rs}$.
Inserting Eq. (\ref{eq:f-rs-pol}) into Eq. (\ref{eq:isz1-sz2}) gives
\begin{eqnarray}
I^{\alpha\beta}({\bf p},{\bf p}^{\prime}) & = & \text{Tr}\left[\Gamma^{\beta}v(s_{1},\mathbf{p}^{\prime})\overline{v}(r_{1},\mathbf{p}^{\prime})\Gamma^{\alpha}u(r_{2},\mathbf{p}-\mathbf{p}^{\prime})\overline{u}(s_{2},\mathbf{p}-\mathbf{p}^{\prime})\right]\nonumber \\
 &  & \times\frac{1}{2}f_{\overline{q}}(x,\mathbf{p}^{\prime})\left[\delta_{r_{1}s_{1}}-P_{\mu}^{\overline{q}}(x,\mathbf{p}^{\prime})n_{j}^{(-)\mu}(-\mathbf{p}^{\prime})\tau_{r_{1}s_{1}}^{j}\right]\nonumber \\
 &  & \times\frac{1}{2}f_{q}(x,\mathbf{p}-\mathbf{p}^{\prime})\left[\delta_{r_{2}s_{2}}-P_{\mu}^{q}(x,\mathbf{p}-\mathbf{p}^{\prime})n_{j}^{(+)\mu}(\mathbf{p}-\mathbf{p}^{\prime})\tau_{r_{2}s_{2}}^{j}\right]\nonumber \\
 & = & \frac{1}{4}f_{\overline{q}}(x,\mathbf{p}^{\prime})f_{q}(x,\mathbf{p}-\mathbf{p}^{\prime})\nonumber \\
 &  & \times\text{Tr}\left\{ \Gamma^{\beta}\left(p^{\prime}\cdot\gamma-m_{\overline{q}}\right)\left[1+\gamma_{5}\gamma\cdot P^{\overline{q}}(x,\mathbf{p}^{\prime})\right]\Gamma^{\alpha}\right.\nonumber \\
 &  & \times\left.\left[(p-p^{\prime})\cdot\gamma+m_{q}\right]\left[1+\gamma_{5}\gamma\cdot P^{q}(x,\mathbf{p}-\mathbf{p}^{\prime})\right]\right\} ,\label{eq:i-ab}
\end{eqnarray}
where we have used 
\begin{eqnarray}
 &  & v(s_{1},\mathbf{p}^{\prime})\overline{v}(r_{1},\mathbf{p}^{\prime})\left[\delta_{r_{1}s_{1}}-P_{\mu}^{\overline{q}}(x,\mathbf{p}^{\prime})n_{j}^{(-)\mu}(-\mathbf{p}^{\prime})\tau_{r_{1}s_{1}}^{j}\right]\nonumber \\
 & = & \left(p^{\prime}\cdot\gamma-m_{\overline{q}}\right)\left[1+\gamma_{5}\gamma\cdot P^{\overline{q}}(x,\mathbf{p}^{\prime})\right],\label{eq:vv-pol}
\end{eqnarray}
and 
\begin{eqnarray}
 &  & u(r_{2},\mathbf{p}-\mathbf{p}^{\prime})\overline{u}(s_{2},\mathbf{p}-\mathbf{p}^{\prime})\nonumber \\
 &  & \times\left[\delta_{r_{2}s_{2}}-P_{\mu}^{q}(x,\mathbf{p}-\mathbf{p}^{\prime})n_{j}^{(+)\mu}(\mathbf{p}-\mathbf{p}^{\prime})\tau_{r_{2}s_{2}}^{j}\right]\nonumber \\
 & = & \left[(p-p^{\prime})\cdot\gamma+m_{q}\right]\left[1+\gamma^{5}\gamma\cdot P^{q}(x,\mathbf{p}-\mathbf{p}^{\prime})\right].\label{eq:uu-pol}
\end{eqnarray}
In deriving (\ref{eq:vv-pol}) and (\ref{eq:uu-pol}) we have used
\begin{eqnarray}
\frac{1}{2}\epsilon^{\alpha\beta\mu\nu}\sigma_{\mu\nu} & = & \gamma_{5}\gamma^{\alpha}\gamma^{\beta}-g^{\alpha\beta}\gamma_{5},\nonumber \\
n_{j}^{(-)\mu}(-\mathbf{p}^{\prime}) & = & n^{\mu}(\mathbf{n}_{j},\mathbf{p}^{\prime},m_{\overline{q}}),\nonumber \\
p^{\prime\mu}P_{\mu}^{\overline{q}}(x,\mathbf{p}^{\prime}) & = & (p-p^{\prime})^{\mu}P_{\mu}^{q}(x,\mathbf{p}-\mathbf{p}^{\prime})=0.
\end{eqnarray}
Inserting (\ref{eq:i-ab}) into (\ref{eq:i-l1l2}) we obtain 
\begin{eqnarray}
I_{\lambda_{1}\lambda_{2}} & = & \epsilon_{\alpha}^{\ast}(\lambda_{1},{\bf p})\epsilon_{\beta}(\lambda_{2},{\bf p})\nonumber \\
 &  & \times\text{Tr}\left\{ \Gamma^{\beta}\left(p^{\prime}\cdot\gamma-m_{\overline{q}}\right)\left[1+\gamma_{5}\gamma\cdot P_{\overline{q}}(x,\mathbf{p}^{\prime})\right]\Gamma^{\alpha}\right.\nonumber \\
 &  & \times\left.\left[(p-p^{\prime})\cdot\gamma+m_{q}\right]\left[1+\gamma_{5}\gamma\cdot P_{q}(x,\mathbf{p}-\mathbf{p}^{\prime})\right]\right\} .\label{eq:i-lambda1-lambda2}
\end{eqnarray}
From Eq. (\ref{eq:i-lambda1-lambda2}) one arrives at Eq. \ref{eq:spin-density-sz1-sz2}.
Using (\ref{eq:qq-bar-vertex}), the trace in (\ref{eq:i-lambda1-lambda2})
can be worked out and the result of $I_{\lambda_{1}\lambda_{2}}$
is 
\begin{eqnarray}
I_{\lambda_{1}\lambda_{2}} & = & -4g_{V}^{2}B^{2}(\mathbf{p}-\mathbf{p}^{\prime},\mathbf{p}^{\prime})\epsilon_{\alpha}^{\ast}(\lambda_{1})\epsilon_{\beta}(\lambda_{2})\nonumber \\
 &  & \times\left\{ \left(p^{\prime\alpha}P_{\overline{q}}^{\beta}+p^{\prime\beta}P_{\overline{q}}^{\alpha}\right)(p^{\prime}\cdot P_{q})-\left(p^{\prime\alpha}P_{q}^{\beta}+p^{\prime\beta}P_{q}^{\alpha}\right)(p\cdot P_{\overline{q}})\right.\nonumber \\
 &  & +2p^{\prime\alpha}p^{\prime\beta}(1-P_{\overline{q}}\cdot P_{q})+g^{\alpha\beta}\left[p^{\prime}\cdot p+(p^{\prime}\cdot P_{q})(p\cdot P_{\overline{q}})\right]\nonumber \\
 &  & +\left[(m_{q}-m_{\overline{q}})m_{\overline{q}}+p\cdot p^{\prime}\right]\left(P_{\overline{q}}^{\alpha}P_{q}^{\beta}+P_{q}^{\alpha}P_{\overline{q}}^{\beta}-g^{\alpha\beta}P_{\overline{q}}\cdot P_{q}\right)\nonumber \\
 &  & -(m_{\overline{q}}-m_{q})m_{\overline{q}}g^{\alpha\beta}-i(m_{q}-m_{\overline{q}})\varepsilon^{\alpha\beta\mu\nu}p_{\mu}^{\prime}(P_{\nu}^{q}+P_{\nu}^{\overline{q}})\nonumber \\
 &  & \left.-im_{\overline{q}}\varepsilon^{\alpha\beta\mu\nu}p_{\mu}(P_{\nu}^{q}+P_{\nu}^{\overline{q}})\right\} ,\label{eq:i-sz1-sz2}
\end{eqnarray}
where we have used shorthand notations $\epsilon(\lambda)\equiv\epsilon(\lambda,\mathbf{p})$,
$P_{q}\equiv P_{q}(x,\mathbf{p}-\mathbf{p}^{\prime})$, and $P_{\overline{q}}\equiv P_{\overline{q}}(x,\mathbf{p}^{\prime})$.
We can take the sum of $I_{\lambda\lambda}$ over $\lambda$ as 
\begin{eqnarray}
\sum_{\lambda}I_{\lambda\lambda} & = & 2g_{V}^{2}B^{2}(\mathbf{p}-\mathbf{p}^{\prime},\mathbf{p}^{\prime})\nonumber \\
 &  & \times\left\{ \left[-\left(m_{q}+m_{\overline{q}}\right)^{2}+\frac{1}{m_{V}^{2}}\left(m_{\overline{q}}^{2}-m_{q}^{2}\right)^{2}\right](P_{\overline{q}}\cdot P_{q})\right.\nonumber \\
 &  & +\frac{2}{m_{V}^{2}}\left(m_{q}-m_{\overline{q}}\right)^{2}(p\cdot P_{q})(p\cdot P_{\overline{q}})\nonumber \\
 &  & \left.+2m_{V}^{2}+6m_{q}m_{\overline{q}}-m_{\overline{q}}^{2}-m_{q}^{2}-\frac{1}{m_{V}^{2}}\left(m_{\overline{q}}^{2}-m_{q}^{2}\right)^{2}\right\} .\label{eq:trace}
\end{eqnarray}
From Eq. (\ref{eq:trace}) one can obtain $\mathrm{Tr}(\rho_{V})$
in Eq. (\ref{eq:norm-rho00}).

\section{Spin density matrix in nonrelativistic limit}

\label{sec:non-relativistic-limit}We consider $m_{q}=m_{\overline{q}}$
and assume $m_{V}\approx2m_{q}$. In non-relativistic limit, we can
approximate 
\begin{eqnarray}
p^{\mu} & \approx & (m_{V},{\bf 0}),\nonumber \\
p^{\prime\mu} & \approx & (m_{\overline{q}},{\bf 0})\approx(m_{q},{\bf 0}),\nonumber \\
p^{\mu}-p^{\prime\mu} & \approx & (m_{V}-m_{\overline{q}},0)\approx(m_{q},0),\nonumber \\
P_{\overline{q}}^{\mu}(x,\mathbf{p}^{\prime}) & \approx & (0,\mathbf{P}_{\overline{q}}(x,\mathbf{p}^{\prime})),\nonumber \\
P_{q}^{\mu}(x,\mathbf{p}-\mathbf{p}^{\prime}) & \approx & (0,\mathbf{P}_{q}(x,\mathbf{p}-\mathbf{p}^{\prime})),\nonumber \\
\epsilon^{\mu}(\lambda_{1}) & \approx & (0,\boldsymbol{\epsilon}(\lambda_{1})),\nonumber \\
\epsilon^{\mu}(\lambda_{2}) & \approx & (0,\boldsymbol{\epsilon}(\lambda_{2})),\label{eq:non-rel-1}
\end{eqnarray}
which leads to 
\begin{eqnarray}
p^{\prime}\cdot\epsilon^{\ast}(\lambda_{1}) & \approx & 0,\nonumber \\
p^{\prime}\cdot\epsilon(\lambda_{2}) & \approx & 0,\nonumber \\
p^{\prime}\cdot p & \approx & m_{V}m_{q},\nonumber \\
\epsilon^{\ast}(\lambda_{1})\cdot\epsilon(\lambda_{2}) & \approx & -\boldsymbol{\epsilon}^{\ast}(\lambda_{1})\cdot\boldsymbol{\epsilon}(\lambda_{2}).\label{eq:non-rel-2}
\end{eqnarray}
In Eqs. (\ref{eq:non-rel-1}) and (\ref{eq:non-rel-2}) we have used
the shorthand notation $\epsilon(\lambda)\equiv\epsilon(\lambda,\mathbf{p})$.
Using (\ref{eq:non-rel-1}) and (\ref{eq:non-rel-2}), $I_{\lambda_{1}\lambda_{2}}$
in Eq. (\ref{eq:i-sz1-sz2}) has a simple form  
\begin{eqnarray}
I_{\lambda_{1}\lambda_{2}} & = & 4g_{V}^{2}m_{V}m_{q}\left\{ \boldsymbol{\epsilon}^{\ast}(\lambda_{1})\cdot\boldsymbol{\epsilon}(\lambda_{2})\left(1+\mathbf{P}_{\overline{q}}\cdot\mathbf{P}_{q}\right)\right.\nonumber \\
 &  & -[\mathbf{P}_{\overline{q}}\cdot\boldsymbol{\epsilon}^{\ast}(\lambda_{1})][\mathbf{P}_{q}\cdot\boldsymbol{\epsilon}(\lambda_{2})]-[\mathbf{P}_{q}\cdot\boldsymbol{\epsilon}^{\ast}(\lambda_{1})][\mathbf{P}_{\overline{q}}\cdot\boldsymbol{\epsilon}(\lambda_{2})]\nonumber \\
 &  & \left.-i[\boldsymbol{\epsilon}^{\ast}(\lambda_{1})\times\boldsymbol{\epsilon}(\lambda_{2})]\cdot(\mathbf{P}_{q}+\mathbf{P}_{\overline{q}})\right\} .
\end{eqnarray}
One can verify $I_{\lambda_{2}\lambda_{1}}=I_{\lambda_{1}\lambda_{2}}^{*}$.

From (\ref{eq:spin-density-sz1-sz2}), the spin density matrix for
the vector meson in the non-relativistic limit is given by 
\begin{eqnarray}
\rho_{\lambda_{1}\lambda_{2}}^{V}(x,{\bf p}) & = & \frac{\Delta t}{8}g_{V}^{2}m_{V}m_{q}\int\frac{d^{3}\mathbf{p}^{\prime}}{(2\pi\hbar)^{3}}\frac{1}{E_{p^{\prime}}^{\overline{q}}E_{{\bf p}-{\bf p}^{\prime}}^{q}E_{p}^{V}}\nonumber \\
 &  & \times f_{\overline{q}}(x,\mathbf{p}^{\prime})f_{q}(x,\mathbf{p}-\mathbf{p}^{\prime})2\pi\hbar\delta\left(E_{p}^{V}-E_{p^{\prime}}^{\overline{q}}-E_{{\bf p}-{\bf p}^{\prime}}^{q}\right)\nonumber \\
 &  & \times\left\{ \boldsymbol{\epsilon}^{\ast}(\lambda_{1})\cdot\boldsymbol{\epsilon}(\lambda_{2})\left[1+{\bf P}_{q}(x,{\bf p}-{\bf p}^{\prime})\cdot{\bf P}_{\overline{q}}(x,{\bf p}^{\prime})\right]\right.\nonumber \\
 &  & -\left[\mathbf{P}_{q}(x,{\bf p}-{\bf p}^{\prime})\cdot\boldsymbol{\epsilon}(\lambda_{2})\right]\left[\mathbf{P}_{\overline{q}}(x,{\bf p}^{\prime})\cdot\boldsymbol{\epsilon}^{\ast}(\lambda_{1})\right]\nonumber \\
 &  & -\left[\mathbf{P}_{q}(x,{\bf p}-{\bf p}^{\prime})\cdot\boldsymbol{\epsilon}^{\ast}(\lambda_{1})\right]\left[\mathbf{P}_{\overline{q}}(x,{\bf p}^{\prime})\cdot\boldsymbol{\epsilon}(\lambda_{2})\right]\nonumber \\
 &  & \left.-i\left[\boldsymbol{\epsilon}^{\ast}(\lambda_{1})\times\boldsymbol{\epsilon}(\lambda_{2})\right]\cdot\left[\mathbf{P}_{q}(x,{\bf p}-{\bf p}^{\prime})+\mathbf{P}_{\overline{q}}(x,{\bf p}^{\prime})\right]\right\} .\label{eq:spin-density-matrix}
\end{eqnarray}
We can simplify the above formula by using the shorthand notation
\begin{eqnarray}
Dp^{\prime} & \equiv & \frac{\Delta t}{8}g_{V}^{2}m_{V}m_{q}\int\frac{d^{3}\mathbf{p}^{\prime}}{(2\pi\hbar)^{3}}\frac{1}{E_{p^{\prime}}^{\overline{q}}E_{{\bf p}-{\bf p}^{\prime}}^{q}E_{p}^{V}}\nonumber \\
 &  & \times f_{\overline{q}}(x,\mathbf{p}^{\prime})f_{q}(x,\mathbf{p}-\mathbf{p}^{\prime})2\pi\hbar\delta\left(E_{p}^{V}-E_{p^{\prime}}^{\overline{q}}-E_{{\bf p}-{\bf p}^{\prime}}^{q}\right).\label{eq:weight-integral}
\end{eqnarray}
We can put $\rho_{\lambda_{1}\lambda_{2}}^{V}$ into a matrix form
\begin{eqnarray}
\rho^{V} & = & \left(\begin{array}{ccc}
\rho_{11} & \rho_{10} & \rho_{1,-1}\\
\rho_{1,0}^{*} & \rho_{00} & \rho_{0,-1}\\
\rho_{1,-1}^{*} & \rho_{0,-1}^{*} & \rho_{-1,-1}
\end{array}\right).
\end{eqnarray}
Note that $\rho^{V}$ is a Hermitian matrix and we have suppressed
the index 'V' in all elements.

For a given spin quantization direction ${\bf n}_{3}$, we can construct
$\boldsymbol{\epsilon}(\lambda)$ as follows
\begin{eqnarray}
\boldsymbol{\epsilon}(0) & = & {\bf n}_{3},\nonumber \\
\boldsymbol{\epsilon}(1) & = & -\frac{1}{\sqrt{2}}\left({\bf n}_{1}+i{\bf n}_{2}\right),\nonumber \\
\boldsymbol{\epsilon}(-1) & = & \frac{1}{\sqrt{2}}\left({\bf n}_{1}-i{\bf n}_{2}\right),
\end{eqnarray}
where ${\bf n}_{1}$, ${\bf n}_{2}$ and ${\bf n}_{3}$ form orthogonal
basis vectors in the rest frame of the vector meson. From (\ref{eq:spin-density-matrix})
we obtain 
\begin{eqnarray}
\rho_{11} & = & \int Dp^{\prime}\left(1+{\bf n}_{3}\cdot{\bf P}_{q}\right)\left(1+{\bf n}_{3}\cdot{\bf P}_{\overline{q}}\right),\nonumber \\
\rho_{10} & = & \frac{1}{\sqrt{2}}\int Dp^{\prime}\left\{ \left[({\bf n}_{1}-i{\bf n}_{2})\cdot{\bf P}_{q}\right]\left(1+{\bf n}_{3}\cdot{\bf P}_{\overline{q}}\right)\right.\nonumber \\
 &  & \left.+\left[({\bf n}_{1}-i{\bf n}_{2})\cdot{\bf P}_{\overline{q}}\right]\left(1+{\bf n}_{3}\cdot{\bf P}_{q}\right)\right\} ,\nonumber \\
\rho_{1,-1} & = & \int Dp^{\prime}\left[({\bf n}_{1}-i{\bf n}_{2})\cdot{\bf P}_{q}\right]\left[({\bf n}_{1}-i{\bf n}_{2})\cdot{\bf P}_{\overline{q}}\right],\nonumber \\
\rho_{00} & = & \int Dp^{\prime}\left\{ 1+{\bf P}_{q}\cdot{\bf P}_{\overline{q}}-2\left({\bf n}_{3}\cdot{\bf P}_{q}\right)\left({\bf n}_{3}\cdot{\bf P}_{\overline{q}}\right)\right\} \nonumber \\
\rho_{-1,0} & = & -\frac{1}{\sqrt{2}}\int Dp^{\prime}\left\{ \left[({\bf n}_{1}+i{\bf n}_{2})\cdot{\bf P}_{q}\right]\left(1-{\bf n}_{3}\cdot{\bf P}_{\overline{q}}\right)\right.\nonumber \\
 &  & \left.+\left[({\bf n}_{1}+i{\bf n}_{2})\cdot{\bf P}_{\overline{q}}\right]\left(1-{\bf n}_{3}\cdot{\bf P}_{q}\right)\right\} ,\nonumber \\
\rho_{-1,-1} & = & \int Dp^{\prime}\left(1-{\bf n}_{3}\cdot{\bf P}_{q}\right)\left(1-{\bf n}_{3}\cdot{\bf P}_{\overline{q}}\right),
\end{eqnarray}
where we have used shorthand notations ${\bf P}_{q}\equiv{\bf P}_{q}(x,{\bf p}-{\bf p}^{\prime})$
and ${\bf P}_{\overline{q}}\equiv{\bf P}_{\overline{q}}(x,{\bf p}^{\prime})$.
The 00-element of the normalized density matrix is given by 
\begin{eqnarray}
\overline{\rho}_{00} & = & \frac{\rho_{00}}{\rho_{11}+\rho_{00}+\rho_{-1,-1}}\nonumber \\
 & = & \frac{\int Dp^{\prime}\left[1+{\bf P}_{q}\cdot{\bf P}_{\overline{q}}-2\left({\bf n}_{3}\cdot{\bf P}_{q}\right)\left({\bf n}_{3}\cdot{\bf P}_{\overline{q}}\right)\right]}{\int Dp^{\prime}\left(3+{\bf P}_{q}\cdot{\bf P}_{\overline{q}}\right)},
\end{eqnarray}
where $N\equiv\int Dp^{\prime}$ is the normalization constant. If
the magnitude of the polarization is much smaller than 1, we can make
a Taylor expansion in it and obtain
\begin{eqnarray}
\overline{\rho}_{00}(x,\mathbf{p}) & \simeq & \frac{1}{3}+\frac{2}{9N}\int Dp^{\prime}\left[\left({\bf n}_{1}\cdot{\bf P}_{q}\right)\left({\bf n}_{1}\cdot{\bf P}_{\overline{q}}\right)+\left({\bf n}_{2}\cdot{\bf P}_{q}\right)\left({\bf n}_{2}\cdot{\bf P}_{\overline{q}}\right)\right.\nonumber \\
 &  & \left.-2\left({\bf n}_{3}\cdot{\bf P}_{q}\right)\left({\bf n}_{3}\cdot{\bf P}_{\overline{q}}\right)\right].
\end{eqnarray}
If we assume ${\bf P}_{q}$ and ${\bf P}_{\overline{q}}$ are only
in the direction of ${\bf n}_{3}$, i.e. 
\begin{equation}
{\bf n}_{x}\cdot{\bf P}_{q}={\bf n}_{y}\cdot{\bf P}_{q}={\bf n}_{x}\cdot{\bf P}_{\overline{q}}={\bf n}_{y}\cdot{\bf P}_{\overline{q}}=0,
\end{equation}
we obtain 
\begin{equation}
\overline{\rho}_{00}(x,\mathbf{p})\approx\frac{1}{3}-\frac{4}{9N}\int Dp^{\prime}\left[{\bf n}_{3}\cdot{\bf P}_{q}(x,{\bf p}-{\bf p}^{\prime})\right]\left[{\bf n}_{3}\cdot{\bf P}_{\overline{q}}(x,{\bf p}^{\prime})\right].
\end{equation}
which recovers the similar form to the previous result but expressed
in terms of the weighted integral $Dp^{\prime}$ in (\ref{eq:weight-integral}).

\bibliographystyle{apsrev}
\bibliography{ref-kb-vm}

\end{document}